\documentclass[amsmath,amssymb,twocolumn,floatfix,aps,showpacs,superscriptaddress,nofootinbib,notitlepage]{revtex4-1}

\usepackage{amsmath,amsthm,amssymb,bm,hyperref,graphicx,color,float,changes,mathtools}
\allowdisplaybreaks

\newcommand{\be}{\begin{equation}}
\newcommand{\ee}{\end{equation}}
\newcommand{\bea}{\begin{eqnarray}}
\newcommand{\eea}{\end{eqnarray}}
\newcommand{\6}{\partial}
\newcommand{\g}{\gamma}

\newcommand{\inti}{\int_{-\infty}^{+\infty}}

\begin{document}

\title{Momentum reconstruction and contact of the one-dimensional Bose-Fermi mixture}

\author{Ovidiu I. P\^{a}\c{t}u}
\affiliation{Institute for Space Sciences, Bucharest-M\u{a}gurele, R 077125, Romania}
\author{Andreas Kl\"umper}
\affiliation{Fakult\"at f\"ur Mathematik und
    Naturwissenschaften, Bergische Universit\"at Wuppertal, 42097 Wuppertal,
    Germany}

\begin{abstract}

We investigate the one-dimensional mixture of scalar bosons and spin polarized
fermions interacting through a $\delta$-function potential. Using a
thermodynamic description derived by employing a lattice embedding of the
continuum model and the quantum transfer matrix method we perform a detailed
analysis of the contact and quantum critical behaviour.  We show that the
compressibility Wilson ratio presents anomalous enhancement at the quantum
critical points and that the boundaries of the quantum critical regions can be
well mapped by the maxima of the specific heat. As a function of the coupling
strength and temperature the contact presents nonmonotonic behavior. In the
strong coupling regime the local minimum exhibited by the contact as a
function of temperature is accompanied by a significant momentum
reconstruction at both low and high momenta. This momentum reconstruction
occurs as the system crosses the boundary between the Tomonaga-Luttinger
liquid phase to the spin-incoherent regime and provides an experimental
signature of the transition.

\end{abstract}

\maketitle

\section{Introduction}\label{s1}

Physical systems of ultracold atomic gases are characterized by a high degree
of control over interaction strength, statistics and dimensionality which
makes them ideal candidates for the investigation of various quantum many-body
phenomena \cite{BDZ,CCGOR,GBL2}. The absence of defects and impurities makes
these systems particularly suited for the simulation of many condensed matter
models but at the same time they also allow for the creation of more exotic
quantum systems. One example is the degenerate mixture of bosons and fermions
which has been experimentally realized in various trap and lattice
geometries. The study of Bose-Fermi mixtures (BFM) is extremely important from
the experimental point of view due to the sympathetic cooling of fermions via
interactions with bosons \cite{Onof} but also theoretically because they
exhibit phases and phenomena which are seldom studied in the condensed matter
context. One-dimensional BFM, which are characterized by enhanced quantum
fluctuations, have been investigated - both on the lattice and the continuum -
using mean-field theory \cite{Das,AVT,MJP,RKBLG}, bosonization (Tomonaga-Luttinger
liquid) \cite{CH, MWHLD,FP,MW,OTS,RPR,Sch2}, density waves \cite{MYS,NY}, exact
solutions \cite{LY1,Lai,Lai1,BBGO,
  OBBG,GBL1,ID1,ID2,HZL,YCZ,Hao1,Hao2,GCCL,YGZC,Schl}, and various numerical
approaches \cite{TM, PTHR,SP,MF,ZBSRDR,BPPS,RI,WHZ,NWB}. The phase diagram is very
rich and contains Mott insulators, spin and charge density waves, phase
separation, Tomonaga-Luttinger and spin-incoherent liquids and Wigner
crystals. In recent years there have also been an increasing number of studies
on few-body mixtures, which are, in general, focused on the strong coupling
regime.  Various methods are employed such as: the multi-component
generalization of the Bose-Fermi mapping \cite{GM,FVMM1,FVGMM,DJARMV1,
  DJARMV2,DAFAMV,LJB, LYZ,CCG,HGC}, approximation by spin-chains \cite{DBBRS,deu1,deu2},
energy-functional techniques \cite{VFJVZ,LKTZ,DBZ,BDZ2} and trial wave
functions \cite{BBF,CSS,SMS}.

In this article we study the one-dimensional (1D) mixture of scalar bosons and
spin polarized fermions with contact interactions in the continuum. This
system has been investigated in several papers but the vast majority of them
were restricted to the study of the ground state.  However, experiments are
performed at finite temperature which highlights the need for the computation
of accurate thermodynamic data. For example, many multi-component systems
present quantum phase transitions (QPTs) at zero temperature \cite{S} as
certain parameters are varied (pressure, magnetic field, doping, etc.). The
effects of these QPTs can also be detected at finite temperature in the
so-called quantum critical (QC) region which is characterized by strong
coupling of the thermal and quantum fluctuations. While the zero temperature
phase diagram gives the quantum critical points the determination of the
boundaries of the QC regions can be done only by computing the thermodynamic
properties.

The 1D BFM with contact interactions is integrable when the masses of the
fermions and bosons and all the coupling strengths are independently equal
\cite{LY1,HZL,ID1,ID2}. In this case powerful methods associated with Bethe
Ansatz \cite{KBI,EFGKK} can be employed to calculate various zero and finite
temperature properties. In particular the thermodynamics of the system can be
derived using the thermodynamic Bethe ansatz (TBA)\cite{YY,T}. In general,
thermodynamic descriptions of integrable models derived using the TBA are
characterized by an infinite number of integral equations \cite{T} which makes
their numerical implementation very difficult.
While the BFM is one of the very few exceptions from this rule \cite{YCZ} the
TBA thermodynamics of a large number of integrable multi-component systems
like the two-component Fermi (2CFG) \cite{Y1,G1} or Bose gas (2CBG) \cite{Y1}
suffer from the same drawback. Other notable exceptions are systems characterized
by $q$-deformed  algebras at special roots of unity which quite typically leads
to a truncation.
One way of circumventing these
  difficulties is provided by the quantum transfer matrix (QTM) method
  \cite{Suz,SI, Koma,SAW,K1,K2} which has the advantage of producing a finite
  number of integral equations that are easier to implement numerically. In
  Refs.~\cite{KP,PK1,PK2} the authors succeeded in deriving such thermodynamic
  descriptions for the 2CBG and 2CFG and in this article we show that the same
  method can also applied in the case of the Bose-Fermi mixture. Our result
  hints strongly that similar efficient thermodynamic descriptions involving
  only $\kappa$ integral equations for a $\kappa$-component system can be
  derived using the same method.

We use this result to perform a detailed analysis of the universal Tan contact
\cite{Tan1,Tan2, Tan3,OD,BP1,BKP,ZL,CAL,WC1,WC2,VZM1,VZM2,BZ,PK3} which
governs the $1/k^4$ of the momentum distribution. At finite temperature and as
a function of the coupling strength the contact presents local maxima for
small values of the boson fraction, a feature which is not present at zero
temperature.  Even more interesting, the contact develops a local minimum as a
function of the temperature which results in a counterintuitive momentum
reconstruction at the system's transition from the TLL phase to the incoherent
regime.
In addition, we determine the boundaries of the quantum critical
regions which can be identified with the maxima of the grand canonical
specific heat. Similar to the case of the 2CBG \cite{PKF} the Wilson ratio
presents anomalous enhancement in the vicinity of the quantum critical points
and can be used to distinguish between different phases.

The plan of the paper is as follows. In Sec.~\ref{S2} we introduce the model
and in Sec.~\ref{S3} we present the TBA thermodynamics and our results derived
in the quantum transfer matrix framework. The analysis of the contact and
momentum reconstruction is presented in Sec.~\ref{S4} and the determination of
the boundaries of the QC regions is performed in Sec.~\ref{S5}. The derivation
of the thermodynamics is outlined in Secs.~\ref{S6} and \ref{S7}. We conclude
in Sec.~\ref{S8}.

\section{The model}\label{S2}

The model investigated in this article describes one-dimensional scalar bosons and spinless
fermions with contact interactions. The Hamiltonian in second quantization is
\begin{align}\label{ham}
H&=\int\,dx \sum_{\sigma\in\{B,F\}}\left(\frac{\hbar^2}{2 m_\sigma} \6_x\Psi^\dagger_\sigma
\6_x\Psi_\sigma-\mu_\sigma \Psi^\dagger_\sigma\Psi_\sigma\right) \nonumber \\
&\ \ \ \ \ +\frac{g_{BB}}{2}\Psi^\dagger_B\Psi^\dagger_B\Psi_B\Psi_B+g_{BF}\Psi^\dagger_B
\Psi^\dagger_F\Psi_F\Psi_B \,
\end{align}
where $\Psi_B(x)$ and $\Psi_F(x)$ are bosonic and fermionic fields satisfying
canonical commutation and anticommutation relations, $m_B\, ,m_F$ are the
masses of the bosonic and fermionic particles, and $\mu_B$ and $\mu_F$ are the
chemical potentials. In (\ref{ham}), $g_{BB}$ and $g_{BF}$ are the Bose-Bose
and Bose-Fermi interaction strengths which can be expressed in terms of the 1D
scattering lengths $a_{BB}$ and $a_{BF}$ via $g_{\sigma\sigma'}=
-\hbar^2/m_{\sigma\sigma'} a_{\sigma\sigma'}$ with $\sigma,
\sigma'\in\{B,F\}\, $ and $m_{\sigma\sigma'}=(m_\sigma+m_{\sigma'})/ m_\sigma
m_{\sigma'}$ the reduced mass.

The Hamiltonian (\ref{ham}) is integrable when the masses $m_B=m_F=m$ and
coupling strengths are equal $g_{BB}=g_{BF}=g$ \cite{LY1,ID1,ID2}. This is the
case that will be considered in the rest of this article and in order to make
contact with the literature we are going to use units of $\hbar=2m=1$ and
introduce $g=2c$ with $c>0$. For a system of $M$ particles of which $M_B$ are
bosons and $M_F=M-M_B$ are fermions the energy spectrum of (\ref{ham}) is
\cite{ID1,ID2}
\be\label{energyBF}
E_{BF}=\sum_{j=1}^M \left(k_j^{(1)}\right)^2-\mu_B M_B-\mu_F(M-M_B)\, ,
\ee
with $\{k_j^{(1)}\}_{j=1}^M$ satisfying the Bethe ansatz equations (BAEs)
\begin{subequations}\label{baeBF}
\begin{align}
e^{i k_s^{(1)} L_{BF}}&=\prod_{p=1}^{M_B}\frac{k_s^{(1)}-k_p^{(2)}+i c/2}{k_s^{(1)}-k_p^{(2)}-i c/2}\, ,\ \ s=1,\cdots,M\, ,\\
1&=\prod_{j=1}^M \frac{k_l^{(2)}-k_j^{(1)}+i c/2}{k_l^{(2)}-k_j^{(1)}-i c/2}\, ,\ \ l=1,\cdots,M_B\, ,
\end{align}
\end{subequations}
where $L_{BF}$ is the length of the system and we have assumed periodic
boundary conditions.

\section{Thermodynamics}\label{S3}

\subsection{TBA result}
From the historical point of view the first method employed to determine the
thermodynamics of an integrable model was the thermodynamic Bethe ansatz
\cite{T} introduced by Yang and Yang in their study of the Lieb-Liniger model
\cite{YY}. In the TBA framework the Bose-Fermi mixture was investigated in
Ref. \cite{YCZ}. Introducing an effective magnetic field and chemical
potential defined by $\mu=(\mu_B+\mu_F)/2$ and $2H=\mu_B-\mu_F$ the grand
canonical potential per length is ($\beta=1/T$)
\be\label{grandTBA}
\phi_{YCZ}(\mu,H,\beta)=-\frac{1}{2\pi \beta}\int_{\mathbb{R}}\ln\left[1+e^{-\beta
\epsilon(k)}\right]\, dk\, ,
\ee
with $\epsilon(k)$ satisfying the following system of non-linear integral equations (NLIEs)
\begin{subequations}
\begin{align*}
\epsilon(k)&=k^2-\mu+H-\beta^{-1}\int_{\mathbb{R}}b_1(k-\lambda)
\ln\left[1+e^{-\beta \varphi(\lambda)}\right]\, d\lambda\, ,\\
\varphi(\lambda)&=-2H-\beta^{-1}\int_{\mathbb{R}}b_1(\lambda-k)
\ln\left[1+e^{-\beta \epsilon(k)}\right]\, dk\, ,
\end{align*}
\end{subequations}
with $b_1(k)=c/[2\pi(c^2/4+k^2)]$. It should be noted that in general the TBA
description of multi-component systems involve an infinite number of
NLIEs. Therefore, it is extremely fortunate that in the case of the BFM we
encounter only two equations which is due to the fact that the Bethe equations
(\ref{baeBF}) have only real solutions. However, in the case of all the other
multi-component systems with contact interactions like the 2CBG and 2CFG and
even a lot of single component systems the Bethe equations have complex
solutions which means that the TBA description is very hard to implement
numerically. A more efficient method which has the advantage of producing only
a finite number of integral equations even for models whose BAEs admit complex
solutions is the quantum transfer matrix (QTM) technique.  Even though the QTM
can be defined only for lattice models this difficulty can be circumvented by
considering a lattice embedding for the continuum model.  In
Refs. \cite{KP,PK1,PK2} the authors employed this method and succeeded in
deriving a system of only two NLIEs characterizing the thermodynamics of the
2CBG and 2CFG. The same method can be used in the case of the Bose-Fermi
mixture as we will show below.

\subsection{Alternative thermodynamic description}

\begin{figure}
\includegraphics[width=\linewidth]{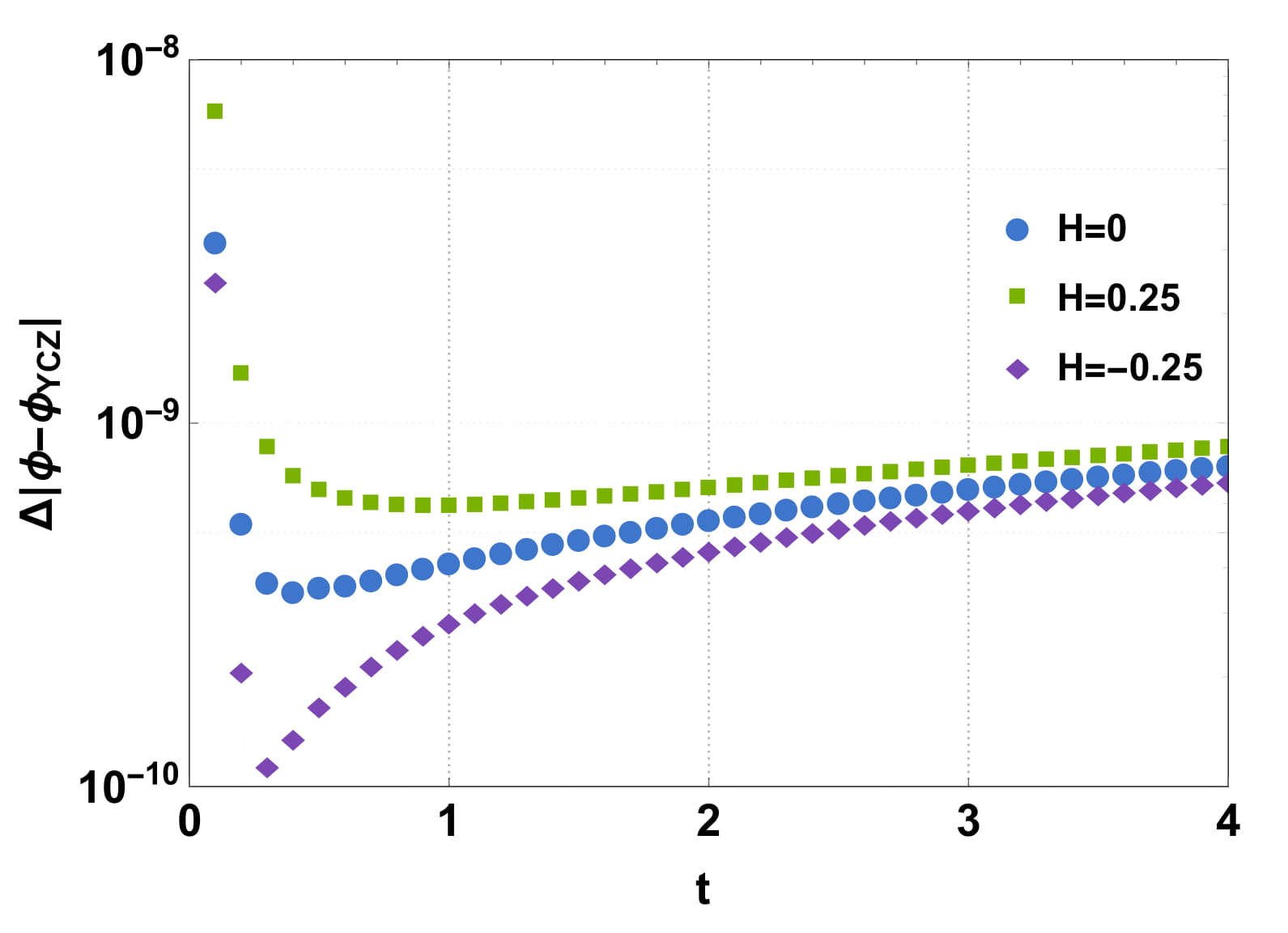}
\caption{Plot of the relative errors between the TBA grand canonical
  potential (\ref{grandTBA}) and our result (\ref{grandQTM}) for $c=1$ and
  $H=-0.25, 0, 0.25 $. Here $t=T/c^2$.  }
\label{error}
\end{figure}
The lattice embedding of the BFM is the Perk-Schultz spin-chain with the
$(-+-)$ grading (see Sec.~\ref{S6}). The derivation of the QTM thermodynamic
description is relatively involved and will be presented in
Sec.~\ref{S7}. Here we present the main result and show the equivalence with
the TBA description. The grand canonical potential per length is
\be\label{grandQTM}
\phi(\mu,H,\beta)=-\frac{1}{2\pi \beta}\int_{\mathbb{R}}\left[\ln A_1(k)+\ln A_2(k)\right]\, dk
\ee
with the two auxiliary functions $a_{1,2}(k)$,  $(A_{1,2}(k)=1+a_{1,2}(k))$ satisfying the
following  system of NLIEs
\begin{subequations}\label{nlieQTM}
\begin{align}
\ln a_1(k)&=-\beta(k^2-\mu-H)+\int_{\mathbb{R}} K_0(k-k') \ln A_1(k')\, dk'\nonumber\\
 & \ \ \ \ \ \ \ \ \ \ +\int_{\mathbb{R}+ i\varepsilon} K_2(k-k') \ln A_2(k')\, dk'\\
 \ln a_2(k)&=-\beta(k^2-\mu+H)\nonumber\\
 & \ \ \ \ \ \ \ \ \ \ + \int_{\mathbb{R}- i\varepsilon} K_1(k-k') \ln A_1(k')\, dk'
\end{align}
\end{subequations}
where $\varepsilon\rightarrow 0$ and the kernels are defined by $ K_0(k)=\frac{1}{2\pi}\frac{2c}
{k^2+c^2}\, , $ $K_1(k)= \frac{1}{2\pi} \frac{c} {k(k+ i c)}\, ,$  and $K_2(k)=\frac{1}{2\pi}\frac{c}{k(k- i c)}\, .$

We can analytically check the validity of our results in some particular cases. In the
noninteracting limit, $c\rightarrow 0,$ using $\lim_{c\rightarrow 0}K_1(k+i\varepsilon)=
\lim_{c\rightarrow 0}
K_2(k-i\varepsilon)
=0$ and $\lim_{c\rightarrow 0}K_2(k)=\delta(k)$ the
NLIEs (\ref{nlieQTM}) decouple
\begin{align*}
\ln a_1(k)&=-\beta(k^2-\mu-H)+\ln\left[1+a_1(k)\right]\, ,\\
\ln a_2(k)&=-\beta(k^2-\mu+H)\, ,
\end{align*}
and can be solved obtaining for the grand canonical potential $\phi(\mu,H,\beta)=\frac{1}
{2\pi\beta} \int_{\mathbb{R}}\ln\left[1-e^{-\beta(k^2-\mu-H)}\right]\, dk - \frac{1}{2\pi\beta}
\int_{\mathbb{R}}\ln\left[1+e^{-\beta(k^2-\mu+H)}\right]\, dk$ which is the known result for
a noninteracting mixture of fermions and bosons with different chemical potentials.  For large
values of $H$ the fermionic degrees of freedom are strongly suppressed, $a_2(k)\sim 0.$
Eqs.~(\ref{nlieQTM}) reduce to the Yang-Yang equation for the Lieb-Liniger model \cite{YY}
\[
\ln a_1(k)=-\beta(k^2-\mu-H)+\int_{\mathbb{R}} K_0(k-k') \ln A_1(k')\, dk'
\]
and $\phi(\mu,H,\beta)=-\frac{1}{2\pi \beta}\int_{\mathbb{R}}\ln \left[1+a_1(k)\right] dk$ which
reproduces the TBA result for single component bosons with contact interactions. In the impenetrable
limit $c\rightarrow \infty$ our result should coincide with the one obtained by Takahashi for
two-component impenetrable fermions
i.e.,
\be\label{imp}
\phi_\infty(\mu,H,\beta)=-\frac{1}{2\pi \beta}\int_{\mathbb{R}}\ln\left[1+2\cosh(\beta H)
e^{-\beta(k^2-\mu)}\right]\, dk\, .
\ee
While we have not succeeded in proving analytically the equivalence of our result with (\ref{imp})
we have checked it numerically and found perfect agreement.

The equivalence of the TBA and QTM thermodynamic descriptions is shown in  Fig.~\ref{error} where
we plot the numerically evaluated relative error defined as
\be\label{rerr}
\Delta|\phi-\phi_{YCZ}|=\frac{|\phi-\phi_{YCZ}|}{\mbox{Max }[\phi,\phi_{YCZ}]}\, ,
\ee
which shows that (\ref{grandTBA}) and (\ref{grandQTM}) (modulo numerical errors) produce identical
results. Because  in both cases we have $\phi(c,\mu, H,T)=c^3 \phi(1,\mu/c^2, H/c^2,T/c^2)$ it is
sufficient to consider only $c=1$.  The computational complexities of both descriptions are the same
which means that choosing one of them is a matter  of personal choice. In the rest of the paper we
use (\ref{grandQTM}) and (\ref{nlieQTM}) mainly because our auxiliary  functions have zero asymptotics
at infinity resulting in a more precise treatment  of convolutions using the Fast Fourier Transform.

The thermodynamic descriptions for the 2CBG \cite{KP,PK1}, 2CFG \cite{PK2} and BFM, (\ref{grandTBA}) and
(\ref{nlieQTM}), derived in the quantum transfer matrix framework involve only two auxiliary functions,
$a_{1,2}(k),$ and the same expression for the grand canonical potential (\ref{grandQTM}). The system of
NLIEs is different in each case and can be  compactly written as ($[f*g](x)=\int_{\mathbb{R}}
f(x-x')g(x')\, dx'$)
\be
\left(\begin{array}{c}\ln a_1(k)\\ \ln a_2(k)\end{array}\right)=
\left(\begin{array}{c} d_1(k)\\ d_2(k)\end{array}\right)+\mathbf{K}*
\left(\begin{array}{c}\ln A_1(k)\\ \ln A_2(k)\end{array}\right)
\ee
with $d_j(k)=-\beta(k^2+\mu+(-1)^j H)$ and kernel matrices
\be
\mathbf{K}_{BB}=\left(\begin{array}{cc} K_0 & K_2 \\ K_1 & K_0\end{array}\right)\, ,\ \ \ \
\mathbf{K}_{FF}=\left(\begin{array}{cc} 0 & K_2 \\ K_1 & 0\end{array}\right)\, ,
\ee
for the 2CBG and 2CFG and
\be
\mathbf{K}_{BF}=\left(\begin{array}{cc} K_0 & K_2 \\ K_1 & 0\end{array}\right)\, ,
\ee
for the Bose-Fermi mixture. It is therefore tempting to conjecture that the thermodynamics of a three-component
system with contact interactions can be described by three auxiliary functions
$a_{i}(k)\, , i=1, 2, 3,\,
A_i(k)=1+a_{i}(k), $  with  grand canonical potential
\[
\phi(\{\mu_i\}_{i=1}^3,\beta)=-\frac{1}{2\pi \beta}\int_{\mathbb{R}}\ln A_1(k)+\ln A_2(k)+\ln A_2(k)\,  dk
\]
and $a_i(k)$ satisfying
\be
\left(\begin{array}{c}\ln a_1(k)\\ \ln a_2(k) \\ \ln a_3(k)\end{array}\right)=
\left(\begin{array}{c} d_1(k)\\ d_2(k) \\ d_3(k)\end{array}\right)+\mathbf{K}*
\left(\begin{array}{c}\ln A_1(k)\\ \ln A_2(k) \\ \ln A_3(k)\end{array}\right)\, ,
\ee
with $d_j(k)=-\beta(k^2+\mu_j).$ In the case of a three-component bosonic and fermionic system
we conjecture that the kernels are
\[
\mathbf{K}_{BBB}=\left(\begin{array}{ccc}
K_0 & K_2 & K_2 \\ K_1 & K_0 & K_2 \\ K_1 & K_1 & K_0\end{array}\right)\, ,\ \ \ \
\mathbf{K}_{FFF}=\left(\begin{array}{ccc}
0 & K_2 & K_2 \\ K_1 & 0 & K_2 \\ K_1 & K_1 & 0\end{array}\right)\, ,\ \ \ \
\]
and in the case of the Bose-Bose-Fermi and Bose-Fermi-Fermi mixtures the kernels are
\[
\mathbf{K}_{BBF}=\left(\begin{array}{ccc}
K_0 & K_2 & K_2 \\ K_1 & K_0 & K_2 \\ K_1 & K_1 & 0\end{array}\right)\, ,\ \ \ \
\mathbf{K}_{BFF}=\left(\begin{array}{ccc}
K_0 & K_2 & K_2 \\ K_1 & 0 & K_2 \\ K_1 & K_1 & 0\end{array}\right)\, .
\]
These conjectured thermodynamic descriptions  present the correct limits when $c\rightarrow 0$ and
when one of the components is suppressed, however, a definitive proof of their validity requires the
numerical checking with the TBA predictions. This will be addressed in a future publication.

\begin{figure*}
\includegraphics[width=\linewidth]{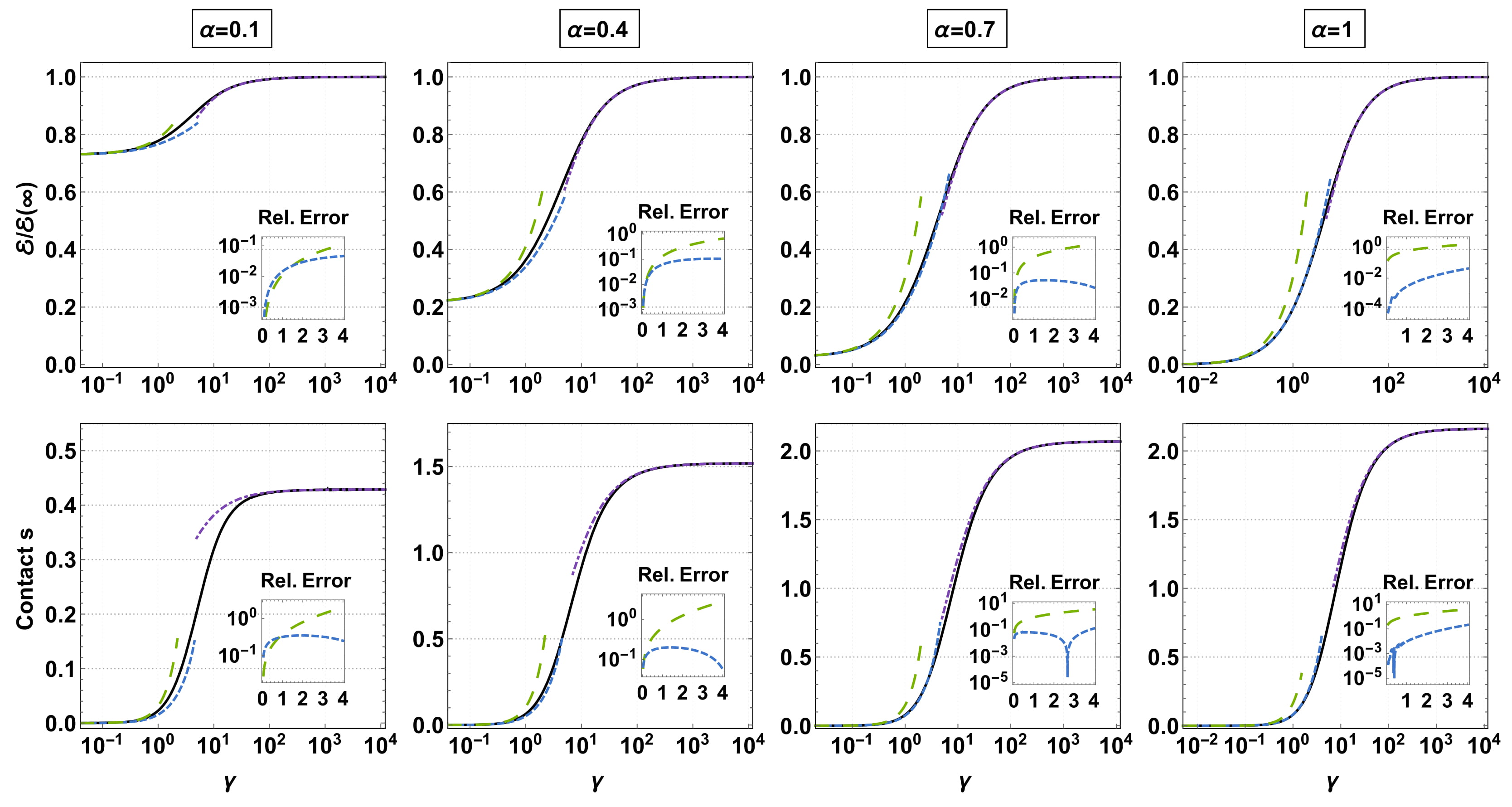}
\caption{{\bf Upper panels.} Energy density normalized by
  $\mathcal{E}(\infty)=n^3\pi^2/3$ (black continuous line) as a function of the
  dimensionless coupling strength $\gamma$ for several values of the boson
  fraction. Also plotted are the strong and weak coupling approximations given
  by Eq.~(\ref{stronge}) (violet dash dotted line),
  Eq.~(\ref{weake}) (long dashed green line) and Eq.~(\ref{weakei}) (short dashed blue line). The
  insets contain the relative errors
  $|\mathcal{E}-\mathcal{E}_{W,WI}|/\mbox{Max }
  [\mathcal{E},\mathcal{E}_{W,WI}]$ of the weak coupling expansions which
  shows that (\ref{weakei}) is an improved approximation. The density is fixed
  $n=1/2$.  {\bf Lower panels.} Normalized total contact $s=\mathcal{C}/(\pi
  n)^4$ as a function of the coupling strength derived from the expressions
  for the energy and approximations using Eq.~(\ref{contact0}).  The insets
  contain the relative errors of the contacts derived from the two weak
  coupling expansions.  }
\label{TanT0}
\end{figure*}

\section{Contact}\label{S4}

The momentum distribution of 1D models with contact interactions present a universal $n(k)\sim
\mathcal{C}/k^4$ decay \cite{OD,BZ,PK3}. The universal coefficient, $\mathcal{C}$, which governs the
asymptotic behavior is called the contact and appears in a series of identities involving macroscopic
properties of the system which are called Tan relations \cite{Tan1,Tan2,Tan3,OD,BP1,BKP,ZL,CAL,WC1,
WC2,VZM1,VZM2,BZ,PK3}. The $1/k^4$ decay and the Tan relations are valid also for nonintegrable systems
in the presence of a trapping potential, at zero or finite temperature and for few- or many-body systems.
For the BFM the  bosonic and fermionic contacts are given by \cite{HGC,PK3}
\begin{align*}
\mathcal{C}_B&=c^2(\langle\Psi^\dagger_B\Psi^\dagger_B\Psi_B\Psi_B\rangle+\langle\Psi^
\dagger_B\Psi^\dagger_F\Psi_F\Psi_B\rangle)\, ,\\
\mathcal{C}_F&=c^2\langle\Psi^\dagger_B\Psi^\dagger_F\Psi_F\Psi_B\rangle\, .
\end{align*}
Even though the individual contacts are hard to compute the total contact can be derived from the
thermodynamics of the system using the Hellmann-Feynman theorem \cite{PK3}
\be\label{contact}
\mathcal{C}=\mathcal{C}_B+\mathcal{C}_F=c^2\left(\frac{\6 \phi}{\6 c}\right)_{\mu,H,T}\, .
\ee

\subsection{Contact at zero temperature}

At zero temperature the thermodynamics of the system is described by a system of Fredholm
integral equations which can be derived from the BAEs (\ref{baeBF}) \cite{ID1,ID2}
\begin{subequations}\label{nlie0}
\begin{align}
\rho_c(k)&=\frac{1}{2\pi}+\int_{-\lambda_0}^{\lambda_0} b_1(k-\lambda)\rho_s(\lambda)\, d\lambda\, ,\\
\rho_s(\lambda)&=\int_{-k_0}^{k_0} b_1(\lambda-k)\rho_c(k)\, dk\, .
\end{align}
\end{subequations}
Here $k_0$ and $\lambda_0$ are two parameters which fix the total density $n=M/L_{BF}$ and the
boson fraction $\alpha=M_B/L_{BF}$ via $n=\int_{-k_0}^{k_0}  \rho_c(k)\, dk, $  and $\alpha=
\int_{-\lambda_0}^{\lambda_0} \rho_s(\lambda)\, d\lambda.$ The energy density of the system is
$\mathcal{E}=\int_{-k_0}^{k_0} k^2 \rho_c(k)\, dk.$ It is useful to introduce the dimensionless
coupling strength $\gamma=c/n.$ The system is in the Tonks-Girardeau regime when $\gamma\gg 1$
and  weakly interacting when $\gamma\ll 1$.

\begin{figure}
\includegraphics[width=\linewidth]{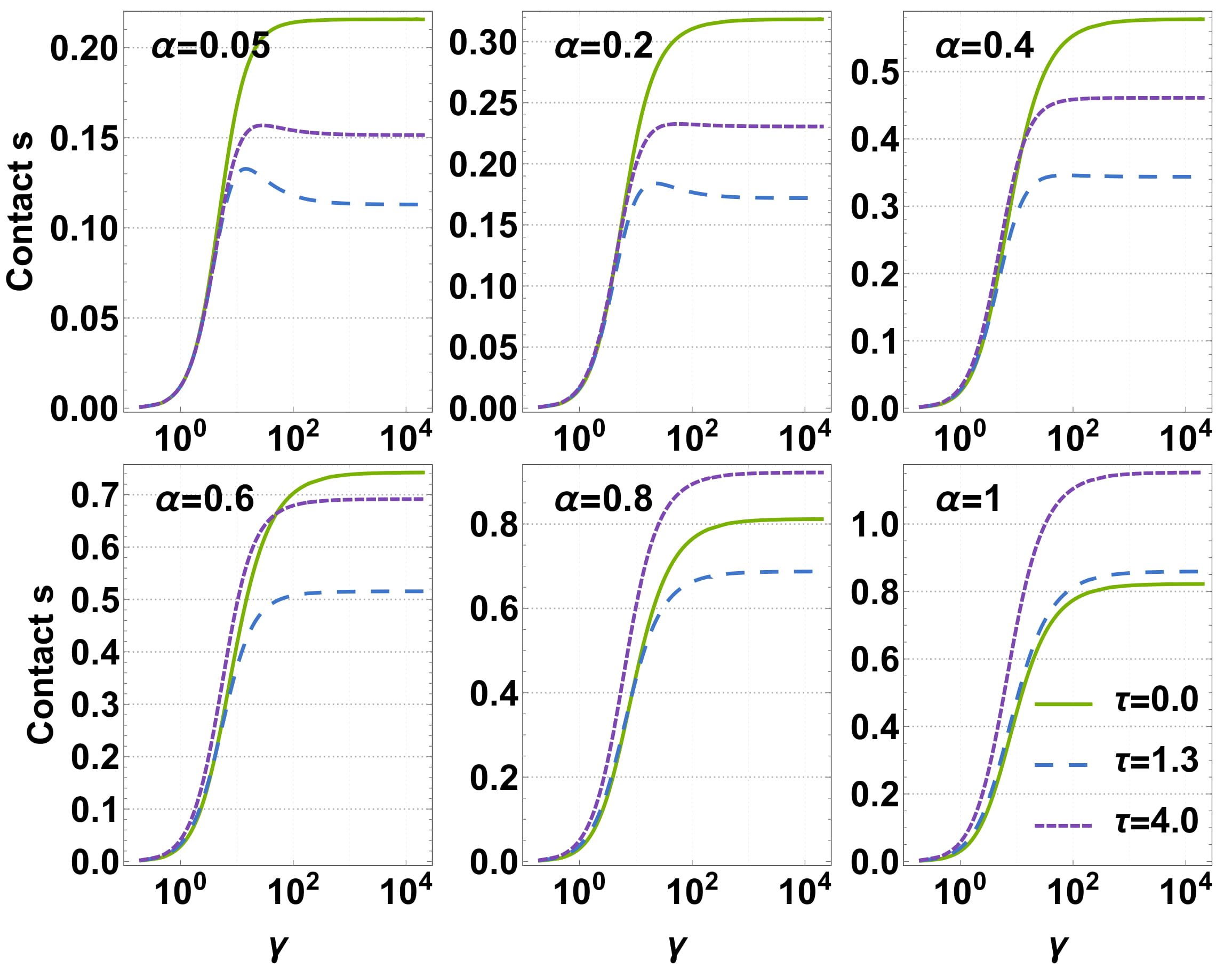}
\caption{Dependence of the dimensionless contact on the coupling strength
  $\gamma$ for
  several values of the reduced temperature ($\tau=T/n^2, n=1/2$) and
  different boson fractions. Compared with the ground state, the contact
  develops a local maximum for small values of $\alpha$, which is more
  pronounced at low but finite temperatures.  }
\label{TanG}
\end{figure}
Once we have computed the energy density the total contact can be derived from Eq.~(\ref{contact})
which at zero temperature takes the form
\be\label{contact0}
\mathcal{C}=n\gamma^2 \left(\frac{\6 \mathcal{E}}{\6 \gamma}\right)_{n,\alpha}\, .
\ee
In general it is relatively easy to derive approximate expressions for the energy in the strong
coupling limit \cite{ID1,ID2,HGC}
\begin{align}\label{stronge}
\mathcal{E}_S(\gamma,\alpha)& \underset{\gamma\gg 1}{\sim} \frac{n^3\pi^2}{3}\left[1-\frac{4}{\gamma}
\left(\alpha+\frac{\sin \pi\alpha}{\pi}\right)\right.\nonumber\\
    & \left.\ \ \ \ \ \ \ \ \ \ \ \ \ \ \ \ \ \ \  + \frac{12}{\gamma^2}\left(\alpha+
\frac{\sin \pi\alpha}{\pi}\right)^2\right]\, ,
\end{align}
however, in the weakly interacting limit serious difficulties are encountered due to the fact
that the $b_1(k)$ kernel becomes a delta function. In this limit only the first term of the
asymptotic expansion was obtained \cite{Das}
\be\label{weake}
\mathcal{E}_W(\gamma,\alpha) \underset{\gamma\ll 1}{\sim} n^3\left[\frac{\pi^2}{3}(1-\alpha)^3+
2\gamma\alpha-\gamma\alpha^2\right]\, .
\ee
One way in which we can improve this approximate expression is to replace the $\gamma$ terms which
are multiplied with powers of the boson fraction with the weak coupling expansion of the Lieb-Liniger
model \cite{LL,Tbose,TW,Prol,LHM}
$
\mathcal{E}_{LL}(\gamma) \underset{\gamma\ll 1}{\sim} \gamma-\frac{4}{3\pi}\gamma^{3/2}+
\left(\frac{1}{6}-\frac{1}{\pi^2}\right)\gamma^2\, ,
$
obtaining
\be\label{weakei}
\mathcal{E}_{WI}(\gamma,\alpha) \underset{\gamma\ll 1}{\sim} n^3\left[\frac{\pi^2}{3}(1-\alpha)^3
+2\mathcal{E}_{LL}(\gamma)\alpha-\mathcal{E}_{LL}(\gamma)\alpha^2\right]\, .
\ee
This expression reduces to the free fermionic result for $\alpha=0$ and reproduces the Lieb-Liniger
expansion when the system is purely bosonic $(\alpha=1)$. In the upper panels of  Fig.~\ref{TanT0}
we present results for the normalized energy density computed using (\ref{nlie0}) together with the
asymptotic expansions at strong and weak coupling. The insets show that (\ref{weakei}) represents a
significant  improvement over (\ref{weake}) and for $\alpha>0.5$ the asymptotic expansions are valid
for almost all values of the coupling strengths. The dimensionless contact $s=\mathcal{C}/(\pi n)^4$
calculated using (\ref{contact0}) is shown in the lower panels of Fig.~\ref{TanT0}. At zero temperature
the contact is a monotonically increasing function of both coupling constant and bosonic fraction.

\subsection{Contact at finite temperature}

\begin{figure}
\includegraphics[width=\linewidth]{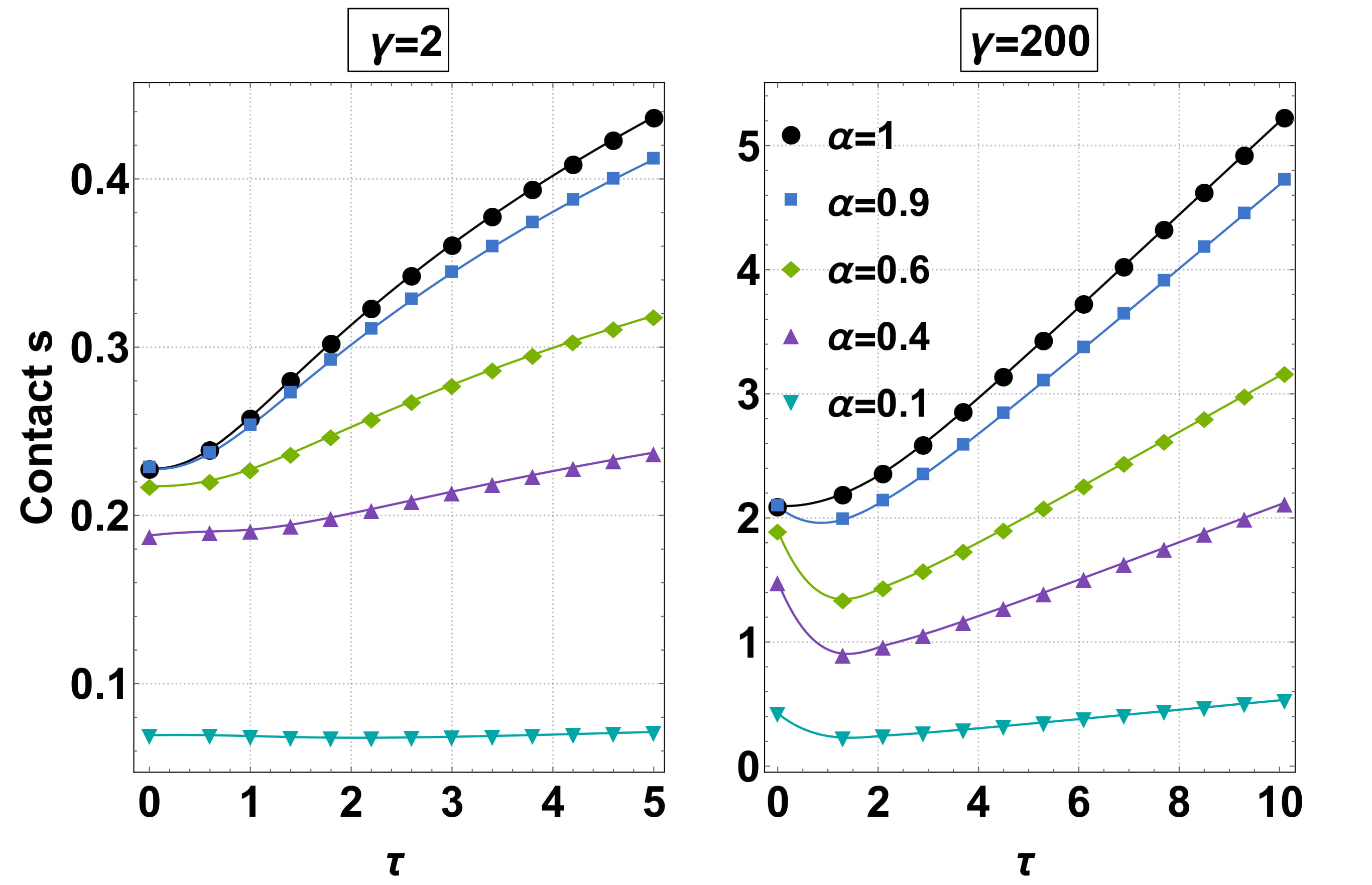}
\caption{Dependence of the dimensionless contact on the reduced temperature for $\gamma=2$ and
$\gamma=200$. At strong coupling the contact presents a pronounced minimum for all values of
the boson fraction except $\alpha=1$.
}
\label{TanT}
\end{figure}

At finite temperature we use (\ref{grandQTM}), (\ref{nlieQTM}) and
(\ref{contact}) for the determination of the contact. The dependence of the
contact on the coupling strength for $\tau=0\, ,1.3\, ,4$ with $\tau=T/n^2$
and different boson fractions is shown in Fig.~\ref{TanG}. We distinguish two
notable features. First, for small values of the boson fraction, $\alpha=0.05$
and $\alpha=0.2$, the contact at finite temperatures develops a local maximum
which is more pronounced at low temperatures. Second, with the exception of
the system close to the purely bosonic case, $\alpha=1$, for large values of
the coupling strength the contact at zero temperature is larger than the one
at finite temperature.
This is rather counterintuitive if we remember that the
contact governs the long tail of the momentum distribution. Therefore, a
smaller contact at higher temperature means that as we increase $T$ the number
of particles with large momenta decreases compared with the ground state. This
phenomenon can be seen more clearly in Fig.~\ref{TanT} where we present the
dependence of the contact on the reduced temperature for moderate and strong
coupling.  For $\gamma=2$ the contact is a monotonically increasing function of
the temperature for all values of the boson fraction, however, at strong
coupling the contact develops a pronounced minimum the only exception being
the case of $\alpha=1$. This momentum reconstruction at low temperatures is a
feature of 1D multi-component systems being present also in the case of the
two-component Fermi \cite{PK2} and Bose \cite{PKF} gas and serves as a
signature of the transition from the Tomonaga-Luttinger liquid phase to the
spin-incoherent regime. In 1D two-component systems there are two relevant
temperature scales \cite{CSZ}: the Fermi temperature $T_F=\pi^2 n^2$ which
characterizes the charge degrees of freedom and $T_0=E_F/\gamma$ which
estimates the bandwidth of the spin excitations (in our case a ``spin excitation''
is represented by the removal of a fermion and the addition of a boson in the
system). In the strong coupling limit we have $0\ll T_0\equiv E_F/\gamma\ll
E_F$ and for $T\in(T_0,T_F)$ the charge degrees of freedom are effectively
frozen while the spin degrees of freedom are highly excited. This regime is
called spin-incoherent \cite{BL,Ber,CZ1,FB,F} and its properties are
significantly different from the more well known Tomonaga-Luttinger liquid
phase. In the BFM the minima of the contact at the transition point out that
the momentum distribution becomes narrower but is also easy to see that this
is also accompanied by significant changes at low momenta. In the TLL regime
the Bose-Bose field correlator presents algebraic decay with
$\langle\Psi_B^\dagger(x)\Psi_B(0)\rangle \sim 1/|x|^{-1/(2 K_b)}$ with
$K_b=1/[(\alpha-1)^2-1]$ derived by Frahm and Palacios \cite{FP} and
numerically confirmed in \cite{ID2}. Therefore, the bosonic momentum
distribution will have a singularity at $k=0$ of the type $n_B(k)\sim
1/|k|^{-1+1/(2K_b)}$. However, in the spin-incoherent regime the correlators
are exponentially decaying which means that the momentum distribution at zero
becomes finite. This shows that there is a significant momentum reconstruction
both at low and large momenta at the transition between the TLL and
spin-incoherent regime.

\section{Boundaries of the quantum critical regions}\label{S5}

In the vicinities of the quantum critical points (QCP) the thermodynamics of
the system is universal and is determined by the universality class of the
quantum phase transition. If we keep the magnetic field fixed and consider the
chemical potential as driving parameter, in the quantum critical region the
pressure can be written as \cite{ZH}
\be\label{scaling}
p(\mu,H,T)\sim p_r(\mu,H)+T^{\frac{d}{z}+1}\mathcal{P}_H\left(\frac{\mu-\mu_c(H)}
{T^{\frac{1}{\nu z}}}\right)\, ,
\ee
with $p_r$ the regular part of the pressure, $d$ the dimension,
$\mathcal{P}_H$ a universal function and $\mu_c(H)$ the quantum critical
point. The universality class of the transition is determined by the
correlation length exponent $\nu$ and the dynamical critical exponent $z$. All
the other thermodynamic quantities can be derived from (\ref{scaling}). For
example, the density and compressibility which are defined by $n=\6 p/\6 \mu$
and $\kappa=\6^2\phi/\6 \mu^2$ are
\begin{align*}
n(\mu,H,T)\sim \frac{\6 p_r}{\6 \mu}(\mu,H)+T^{\frac{d}{z}+1-\frac{1}{\nu z}}\mathcal{P}_H'
\left(\frac{\mu-\mu_c(H)}{T^{\frac{1}{\nu z}}}\right)\, ,\\
\kappa(\mu,H,T)\sim \frac{\6^2 p_r}{\6 \mu^2}(\mu,H)+T^{\frac{d}{z}+1-\frac{2}{\nu z}}
\mathcal{P}_H^{''}\left(\frac{\mu-\mu_c(H)}{T^{\frac{1}{\nu z}}}\right)\, .\\
\end{align*}

We can determine the universality class of the transition by choosing certain values for $z$ and $\nu$
and plotting the scaled pressure $(p-p_r)T^{-\frac{d}{z}-1}$ for several values of temperature \cite{ZH}.
If we have chosen correctly the exponents  all the curves will intersect at the value of the QCP $\mu_c(H)$.
If we plot the scaled pressures as a function of $(\mu-\mu_c(H))/T^{\frac{1}{\nu z}}$ all the curves should
collapse to the  universal curve $\mathcal{P}_H$.
\begin{figure}
\includegraphics[width=\linewidth]{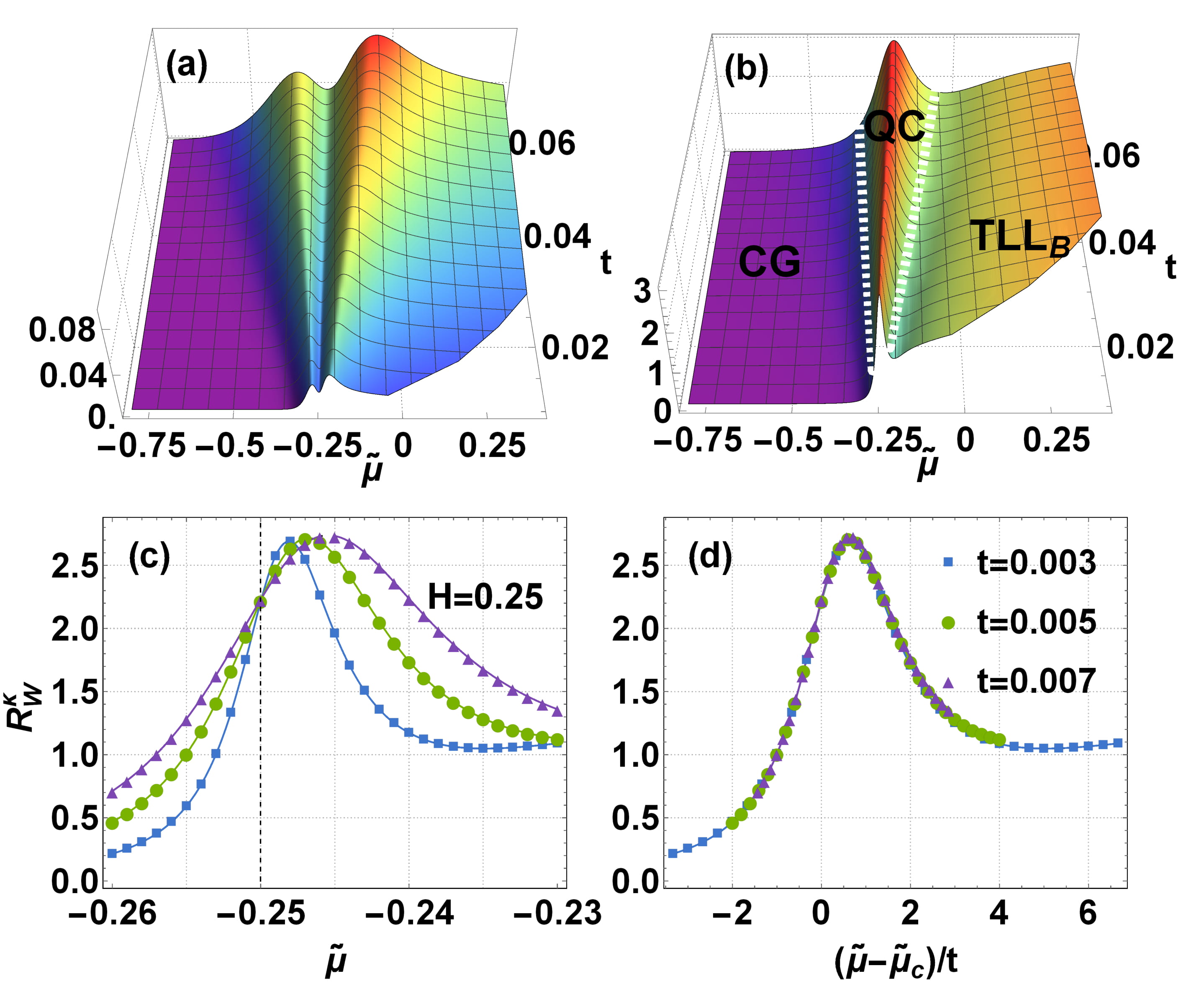}
\caption{{\bf(a)} 3D plot of the grand canonical specific heat for $c=1$ and
  $H=0.25$ as a function of the chemical potential and temperature
  ($\tilde\mu=\mu/c^2 , t=T/c^2$). The lines of local maxima fanning out from
  the QCP, $\tilde{\mu}_c=-H/c^2$, are the boundaries of the QC region.  {\bf
    (b)} 3D plot of the Wilson ratio. The white dashed lines are the
  boundaries of the critical region. CG represents the vacuum (classical gas)
  phase and $TLL_B$ is the Tomonaga-Luttinger liquid phase of single component
  bosons.  {\bf (c)} Plot of the Wilson ratio as a function of the chemical
  potential for three values of temperature. All the curves intersect at the
  QCP (dashed vertical line). The critical exponents are $z=2$ and $\nu=1/2$.
  {\bf (d)} When plotted as a function of $(\tilde\mu-\tilde{\mu}_c(H))/t$ all
  the curves collapse to the universal function $\mathcal{Q}_H$ (see
  Eq.~\ref{wilson}).  }
\label{3D1}
\end{figure}
\begin{figure}
\includegraphics[width=\linewidth]{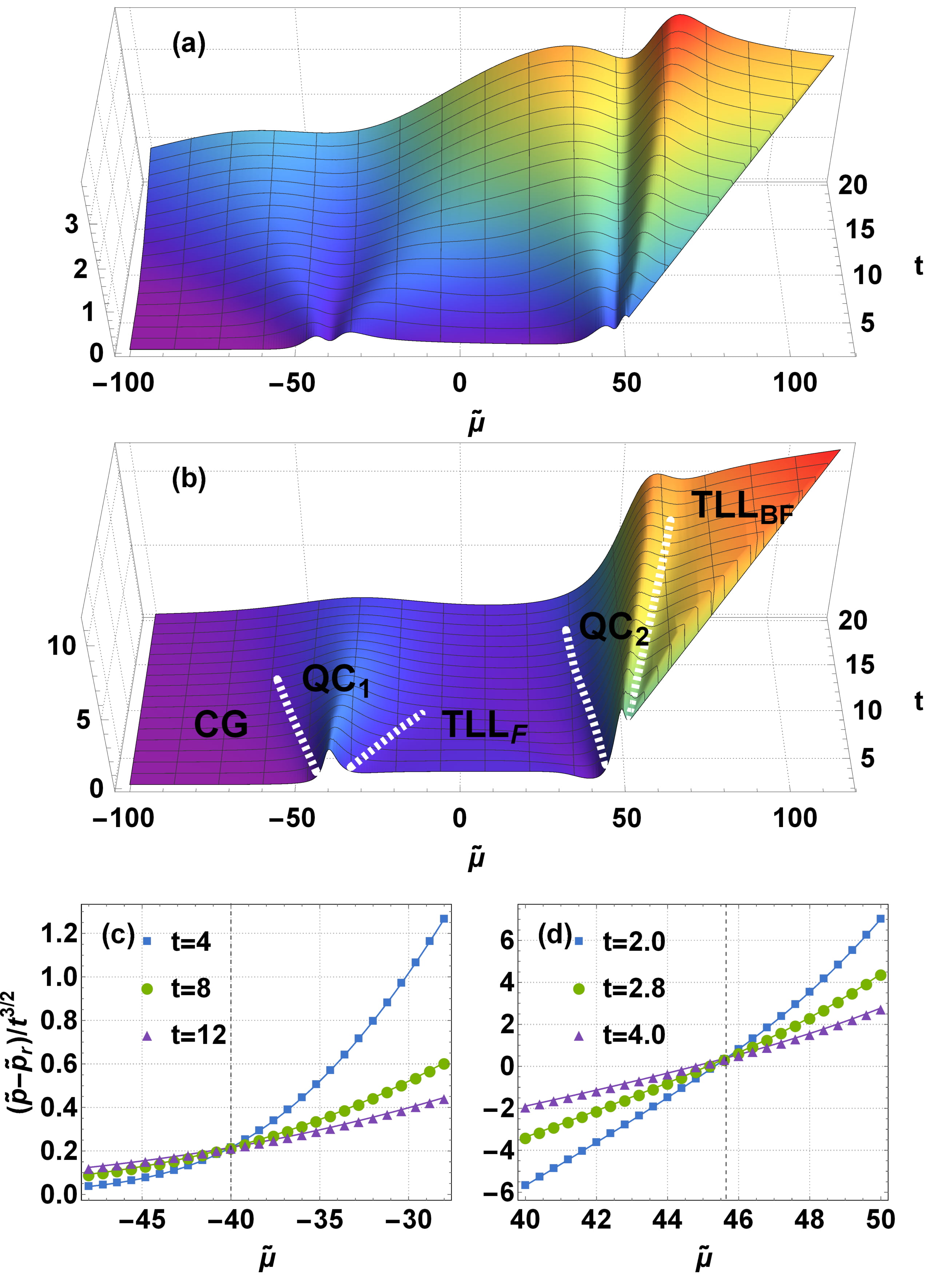}
\caption{{\bf(a)} 3D plot of the grand canonical specific heat for $c=0.05$
  and $H=-0.1$ ($\tilde\mu=\mu/c^2 , t=T/c^2$). In this case we have two sets
  of lines of local maxima which determine the boundaries of the QC regions
  emerging from the quantum critical points situated at
  $\tilde{\mu}_c^{(1)}=-|H|/c^2$, and $\tilde{\mu}_c^{(2)}\sim0.114/c^2$.
  {\bf (b)} 3D plot of the Wilson ratio. The white dashed lines represent the
  boundaries of the critical regions $QC_1$ and $QC_2$. CG, $TLL_F$ and
  $TLL_{BF}$ stand for the classical gas phase, TLL phase of single component
  fermions and TLL phase of bosons and fermions,
  respectively.  {\bf
    (c)} Scaled pressure ($\tilde p=p/c^3$) as a function of the chemical
  potential for three temperatures in the vicinity of the first QCP. For $z=2$
  and $\nu=1/2$ all the curves intersect at $\tilde{\mu}_c^{(1)}=-|H|/c^2$.
  {\bf (d)} Scaled pressure in the vicinity of the second QCP. For $z=2$ and
  $\nu=1/2$ all the curves intersect at $\tilde{\mu}_c^{(2)}\sim 0.114/c^2$.
}
\label{3D2}
\end{figure}

A problem of considerable importance, both theoretically and experimentally, is the determination of
the boundaries of the critical regions. The properties of the system in the CR are fundamentally
different from the ones of  other low-temperature phases and are characterized by the strong coupling
of quantum and thermal fluctuations. In \cite{HJYLG,YCZSG,BGKRT,PKF} it was argued that the grand
canonical specific heat, $c_V=-T\6^2\phi/\6 T^2$ can be used to determine the boundaries of the QC
regions with great precision. This is due to the fact that the grand canonical specific heat is
related to both the energy and number of  particles fluctuations via $k_B T^2 c_v=\langle\delta
(E-\mu N)^2\rangle$ which means that the QC boundaries  can be identified  with the local maxima of
this quantity.  Another important quantity which can be used to identify the  low temperature phases
is the compressibility Wilson ratio \cite{YCLRG,GYFBL,PKF} defined by
\be
R_W^\kappa=\frac{\pi^2k_B^2}{3}T \frac{\kappa}{c_V}\, ,
\ee
with $\kappa$ the compressibility. Because $k_B T\kappa=\langle\delta N^2\rangle$ the Wilson ratio
will be almost constant in the low-temperature phases and will present anomalous enhancement in the
QC regions and will scale like \cite{YCLRG}
\be\label{wilson}
R_W^\kappa\sim \mathcal{Q}_H\left(\frac{\mu-\mu_c(H)}{T^{\frac{1}{\nu z}}}\right)+ w_0 T^{1/2}
\mathcal{F}_H\left(\frac{\mu-\mu_c(H)}{T^{\frac{1}{\nu z}}}\right)\, .
\ee
In the previous equation $\mathcal{Q}_H$ and $\mathcal{F}_H$ are two universal
functions, $w_0$ is a constant and the second term in the right hand side
appears only if $p_r$ is nonzero.

The quantum critical points and the phase diagram at zero temperature were
determined in \cite{YGZC}.
\footnote{It should be noted that the definitions of the chemical potential
  and effective magnetic field employed by us are different from the ones used
  in \cite{YGZC} which will be denoted by the $YGZG$ subscript. We have
  $\mu=\mu_{YGZC}$ and $H=-H_{YGZC}/2$.} The number of QCPs depends on the
sign of the magnetic field. For $H>0$ we have only a QPT from the vacuum to a
single component TLL with critical point $\mu_c=-H$. In Fig.~\ref{3D1} (a) we
present results for the dependence of the grand canonical specific heat on
temperature and chemical potential for $H=0.25$ and coupling strength
$c=1$. The specific heat presents two lines of local maxima fanning out from
the QCP which separate the vacuum (classical gas) and the TLL phase from the
QC region. The Wilson ratio, depicted in Fig.~\ref{3D1} (b), is zero in the
classical gas phase presents a local maximum in the QC region and is slowly
increasing in the TLL phase. In this case $p_r\sim 0$ and $R_W^\kappa$ obeys
the scaling relation (\ref{scaling}) with only the first term on the right
hand side. The scaling and collapse of the curves to the universal function
$\mathcal{Q}_H$ is realized for $z=2$ and $\nu=1/2$ and is presented in
Fig.~\ref{3D1} (c) and Fig.~\ref{3D1} (d).
The value of the critical exponents would seem to point out
that this QPT is in the universality class of free fermions. However it was argued
in \cite{PKF} that in fact this QPT belongs to the universality class of spin-degenerate
impenetrable particle gas with the universal thermodynamics  described by
Takahashi's formula \cite{T}
($x=(\mu+|H|)/T\, ,y=H/T$)
\be
\label{tt1}
p=\frac{T^{3/2}}{2\pi}\inti\ln\left[1+(1+e^{-2|y|})
e^{-k^2+x}\right]\, dk\, ,
\ee
in contrast with the free fermionic case for which ($x'=\mu/T$),
\be\label{tt2}
 p_{FF}=\frac{T^{3/2}}{2\pi}\int\ln\left[\left(1+e^{-k^2+x'+y}\right)
  \left(1+e^{-k^2+x'-y}\right)\right]\, dk\, .
\ee

In the case of fixed negative magnetic field there are two QPTs. The first QCP
is $\mu_c^{(1)}=-|H|$ where the system has a phase transition from the vacuum
to a TLL phase of single component fermions. The value of the second QCP is
determined by ($\tilde{\mu}_c^{(2)}=(\mu_c^{(2)}-H)/c^2$) \cite{YGZC}
\begin{align}
-\frac{2H}{c^2}&=\frac{1}{2\pi}\left[\left(1+4\tilde{\mu}_c^{(2)}\right)
\arctan (4 \tilde{\mu}_c^{(2)})^{1/2}
-(4\tilde{\mu}_c^{(2)})^{1/2} \right]\, ,
\label{mu2H}
\end{align}
where we have a QPT between the single component fermionic TLL to a
two-component TLL composed of fermions and bosons.  The boundaries of the two
QC regions for $c=0.05$ and $H=-0.1$ identified with the maxima of the
specific heat are shown in Fig.~\ref{3D2} (a) and Fig.~\ref{3D2} (b).
In the case of single component systems with QPT belonging to the
free fermionic universality class Maeda et al. \cite{Maed}
derived a universal relation which determines the boundary between the QC and
TLL regions. For $H\gg T$ this relation is also valid for the first QPT of the BFM
due to the fact that in this regime Takahashi's formula (\ref{tt1}) is
equivalent to the pressure of single component free fermions. We stress
that the identification of CR boundaries using the maxima of the specific heat
has the advantage of identifying both boundaries in addition to being valid also for
multi-component systems.

The Wilson ratio presents anomalous enhancement in both critical regions. For
single component systems TLL theory predicts that $R^\kappa_W=K$  \cite{Nin,YCZSG} with
$K$ the  TLL parameter relation which was ``experimentally verified" in the Lieb-Liniger model
\cite{YCZSG}. This identity is also valid for the Bose-Fermi mixture in the TLL regime of
the first QPT for $H\gg T$.

The critical exponents of both QPTs are $z=2\, ,\nu=1/2$
as shown in Fig.~\ref{3D2} (c) and Fig.~\ref{3D2} (d) where the curves for
the scaled pressure at different temperatures intersect at $\mu_c^{(1)}=-|H|$
for the first QPT and at $\mu_c^{(2)}=0.114118 \cdots$ for the second QPT.
While the first transition is in the spin-degenerate universality class characterized by
Eq.~(\ref{tt1}) it is surprising that the second QPT has the same critical
exponents as the free fermionic universality class \cite{Sachdev2}.
We point out that the true universal thermodynamics (\ref{tt1}) in the vicinity of
the critical point $(\mu,H)=(0,0)$ is different from the free spinor fermion
thermodynamics (\ref{tt2}). In the case of the first transition, (\ref{tt1}) and (\ref{tt2}) agree,
for $H\gg T$. For the second critical line it is possible that the universal
thermodynamics is described by a scaling function different from (\ref{tt1}) or (\ref{tt2}).

Lastly, we like to point out certain similarities of the zero
  temperature phase diagram of the Bose-Fermi system with those of the pure
  Bose-Bose and Fermi-Fermi systems with otherwise same mass and interaction
  parameters. For $H\ge 0$ the BF phase diagram is identical to that of the BB
  system with vacuum phase for $\mu<\mu_c$ and completely polarized bosonic
  phase for $\mu>\mu_c$. Viewed from $H>0$, the line $\mu>0$, $H=0$ is a
  transition line into a mixed phase. The location of this line is given by
  the single particle properties of the new admixed particle, the line does
  not depend on its statistics.

For $H< 0$ the BF phase diagram is identical to that of the FF
  system with vacuum phase for $\mu<\mu_c^{(1)}$, completely polarized
fermionic phase for $\mu_c^{(1)}<\mu<\mu_c^{(2)}$, and mixed fermionic-bosonic
  phase for $\mu_c^{(2)}<\mu$. The critical line $\mu_c^{(2)}=\mu_c^{(2)}(H)$
  satisfies (\ref{mu2H}) for the BF and the FF case as can be derived from the
  low temperature limit of the TBA equations for the BF case \cite{YGZC} as
  well as for the FF case \cite{LGSB12}.  When approaching this line from the
  polarized phase, its location is again given by the single particle
  properties of the new admixed particle, the line does not depend on its
  nature.

\section{The Bose-Fermi mixture as the continuum limit of the Perk-Schultz spin chain}\label{S6}

The derivation of the BFM's thermodynamic description, (\ref{grandQTM}) and
(\ref{nlieQTM}), consists of three steps. First, we show that the Perk-Schultz
spin chain \cite{PS1, Sch,BVV,Vega1,VL,Lop} is a lattice embedding of our
continuum model. The thermodynamics of the spin-chain is then investigated
with the quantum transfer matrix technique \cite{Suz,SI, Koma,SAW,K1,K2} which
relates the free energy of the model to the largest eigenvalue of the QTM and
involves only a finite number of NLIEs. Finally, the result for the BFM is
obtained by taking the continuum limit in the lattice result. This method was
first employed in the case of the Lieb-Liniger model \cite{SBGK} and then used
to derive efficient, that is involving only a finite number of NLIEs,
thermodynamic descriptions for the 2CBG \cite{KP,PK1} and 2CFG
\cite{PK2}. Because the ratios of the largest to the next-largest eigenvalues
of the QTM give the correlation lengths of various Green's function the same
algorithm can be used to investigate the asymptotic behavior of correlators in
integrable continuum models \cite{BP,PK4}.

As in the case of the 2CBG and 2CFG the lattice embedding of the Bose-Fermi mixture is the critical $q=3$
Perk-Schultz spin-chain \cite{PS1, Sch,BVV,Vega1,VL,Lop}, the only difference being the grading, which in this
case is  $(-+-)$ (see also \cite{PK1,PK2}). Here, by a lattice embedding we understand a lattice model whose
spectrum and BAEs transform under a suitable scaling limit in the spectrum and BAEs of the continuum model.
The Hamiltonian for an arbitrary grading ($\varepsilon_1,\varepsilon_2,\varepsilon_3),\,
(\varepsilon_i\in\{\pm 1\})$ is
\onecolumngrid
\begin{align}\label{HPS}
\cal{H}_{PS}&=J\varepsilon_1\sum_{j=1}^L\left(\cos\g\sum_{a=1}^3\varepsilon_a\,  e_{aa}^{(j)}e_{aa}^{(j+1)}+
\sum_{\substack{a,b=1\\ a \ne b}}^3 e_{ab}^{(j)}e_{ba}^{(j+1)}
+i\sin\g\sum_{\substack{a,b=1\\ a\ne b}}^3 \mbox{sign}(a-b)e_{aa}^{(j)}e_{bb}^{(j+1)}\right)-\sum_{j=1}^L
\sum_{a=1}^3 h_a e_{aa}^{(j)}\, ,
\end{align}
with $L$ the number of lattice sites, $J>0$ the coupling strength and  $h_1,h_2,h_3$  chemical potentials. Also,
in  (\ref{HPS}) $\g\in[0,\pi]$ is the anisotropy (not to be confused with the dimensionless coupling constant of
the continuum model) and  $e^{(j)}_{ab}=\mathbb{I}_3^{\otimes j-1}\otimes e_{ab}\otimes \mathbb{I}_3^{\otimes L-j}
\, ,$ with  $e_{ab}$ and $\mathbb{I}_3$ the canonical basis and the unit matrix in the space of $3$-by-$3$
matrices. For the  $(-+-)$ grading the energy spectrum is
\be
E_{PS}=\sum_{j=1}^M e_0(v_j^{(1)})+M_1(h_2-h_3)+E_0\, ,\ \ E_0=JL\cos\g-h_1 L\, ,\ \ \
e_0(v)=\frac{J \sin^2\g}{\sin(v-\g)\sin v}\, ,
\ee
with $\{v_s^{(1)}\}_{s=1}^M$ and $\{v_l^{(2)}\}_{l=1}^{M_1}$ satisfying the BAEs
\begin{subequations}\label{baePS}
\begin{align}
\left((-1)\frac{\sin(v_s^{(1)}-\g)}{\sin v_s^{(1)}}\right)^L&=(-1)^{M-1}\prod_{p=1}^{M_1}
\frac{\sin(v_s^{(1)}-v_p^{(2)}-\g)}{\sin(v_s^{(1)}-v_p^{(2)})}\, ,\ \ \ \  s=1\,\cdots,M\, ,\\
\prod_{j=1}^M\frac{\sin(v_l^{(2)}-v_j^{(1)}+\g)}{\sin(v_l^{(2)}-v_j^{(1)})}&=(-1)^{M_1-1}\, ,
\ \ \ \ \ \ \ \ \ \ \ \ \ \ \ \ \ \ \ \ \ \ \ \ \ \ \ \ \ \ \ \ \ \ l=1,\cdots,M_1\, .
\end{align}
\end{subequations}
First, we will show how we can obtain (\ref{baeBF}) from (\ref{baePS}). We consider $v_s^{(1)}\rightarrow i
\delta k_s^{(1)}/\epsilon+\gamma/2$  and  $v_s^{(2)}\rightarrow i\delta k_s^{(2)}/\epsilon+\pi/2$  with
$\epsilon\rightarrow 0$ and  lattice constant $\delta\rightarrow O(\epsilon^2)$. Under this transformation
(\ref{baePS}) become
\begin{align*}
\left((-1)\frac{\sinh(\delta k_s^{(1)}/\epsilon-i\g/2)}{\sinh(\delta k_s^{(1)}/\epsilon+i\g/2)}\right)^L&=
(-1)^{M-1}\prod_{p=1}^{M_1}\frac{\cosh(\delta k_s^{(1)}/\epsilon-\delta k_p^{(2)}/\epsilon-i\g/2)}
{\cosh(\delta k_s^{(1)}/\epsilon-\delta k_p^{(2)}/\epsilon+i\g/2)}\, ,\ \ \ s=1,\cdots,M\, ,\\
\prod_{j=1}^M\frac{\cosh(\delta k_l^{(2)}/\epsilon-\delta k_j^{(1)}/\epsilon-i\g/2)}
{\cosh(\delta k_l^{(2)}/\epsilon-\delta k_j^{(1)}/\epsilon+i\g/2)}&=(-1)^{M_1-1}\, ,\ \ \ \ \ \
\ \ \ \ \ \ \ \ \ \ \ \ \ \ \ \ \ \ \ \ \ \ \ \ \ \ \ \ \ \ \ \ \ \ \ \ \ \ \ \ \  l=1,\cdots,M_1\, .
\end{align*}
In the second step we perform $\g\rightarrow \pi-\epsilon$ with the result
\begin{subequations}\label{ii1}
\begin{align}
\left(\frac{\cosh(\delta k_s^{(1)}+i \epsilon/2)}{\cosh(\delta k_s^{(1)}-i \epsilon/2)}\right)^L&=
(-1)^{M+M_1-1}\prod_{p=1}^{M_1}\frac{\sinh(\delta k_s^{(1)}/\epsilon-\delta k_p^{(2)}/\epsilon+i\epsilon/2)}
{\sinh(\delta k_s^{(1)}/\epsilon-\delta k_p^{(2)}/\epsilon-i\epsilon/2)}\, ,\ \ \ \ s=1,\cdots, M\, ,\\
\prod_{j=1}^M\frac{\sinh(\delta k_l^{(2)}/\epsilon-\delta k_j^{(1)}/\epsilon+i\epsilon/2)}
{\sinh(\delta k_l^{(2)}/\epsilon-\delta k_j^{(1)}/\epsilon-i\epsilon/2)}&=(-1)^{M+M_1-1}\, , \ \ \ \ \ \ \
\ \ \ \ \ \ \ \ \ \ \ \ \ \ \ \ \ \ \ \ \ \ \ \ \ \ \ \ \ \ \ \ \ \ \ \ \ \ \ \ \ l=1,\cdots,M_1\, .
\end{align}
\end{subequations}
\twocolumngrid
Taking the limit $L\rightarrow \infty$  such that $L\delta=L_{BF}$, introducing $c=\epsilon^2/\delta$
and using
\[
\frac{\cosh(\delta k_s^{(1)}+i\epsilon/2)}{\cosh(\delta k_s^{(1)}-i\epsilon/2)}\sim
\frac{1+i\delta k_s^{(1)}/2}{1-i\delta k_s^{(1)}/2}\, ,
\]
we see that Eqs.~(\ref{ii1}) transform into the BAEs of the mixture (\ref{baeBF}) for $M_1+M-1$
even and identifying $M_1=M_B$.
Under the same set of transformations we have
\begin{align*}
E_{PS}-E_0=&\sum_{j=1}^M\left[J\delta^2 \left(k_j^{(1)}\right)^2-J\epsilon^2-J\epsilon^4/4+h_1-h_2\right]\\
&\ \ \ \ \ \ \ \ \ \ \ \ \ \ \ \ \ \ \  +(h_2-h_3)M_1+O(\epsilon^6)\, .
\end{align*}
However, we are interested in the thermodynamical behavior and therefore we can also scale the temperature
in the models in order to have $\beta(E_{PS}-E_0)\rightarrow\bar\beta E_{BF}$ with $E_{BF}$ given by
(\ref{energyBF}). If we consider $J=1$, $\beta=\bar\beta/\delta^2$, $h_1\rightarrow O(\epsilon^2)$ such that
$(J\epsilon^2-h_1)/\delta^2$ is finite and $h_2,h_3\rightarrow O(\epsilon^4)$, we obtain $\beta(E_{PS}-E_0)
\rightarrow\bar \beta E_{BF}$ with $\mu_F=J\epsilon^2+J\epsilon^4/4-h_1+h_2)/\delta^2$ and $\mu_B-\mu_F=
(h_3-h_2)/\delta^2$. The scaling limit presented in this section is the same as the one used in the 2CBG
and 2CFG case (see Table I  of \cite{PK1}) and shows that the thermodynamic behavior of the mixture at all
temperatures can be derived from the low temperature thermodynamics of the lattice model.

\section{Derivation of the thermodynamics for the Perk-Schultz spin-chain}\label{S7}

The free energy of the Perk-Schultz spin-chain can be obtained from the largest eigenvalue of the QTM
as $f(h_1,h_2,h_3,\beta)=-\ln \Lambda_0(0)/\beta$. For a given Trotter number, denoted by $N$, the largest
eigenvalue of the QTM lies in the $(N/2,N/2)$ sector (see Appendix A of \cite{PK1} or \cite{Go1,Go2,Ra1,Ra2}) and can be written as
\be
\Lambda_0(v)=\lambda_1(v)+\lambda_2(v)+\lambda_3(v)\, ,
\ee
with
\be\label{deflambda}
\lambda_j(v)=\phi_-(v)\phi_+(v)\frac{q_{j-1}(v-i\tilde{\epsilon}_j\g)}{q_{j-1}(v)}\frac{q_{j}
(v+i\tilde{\epsilon}_j\g)}{q_{j}(v)}e^{\beta\tilde{h}_j}\, ,
\ee
where $(\tilde{\epsilon}_1,\tilde{\epsilon}_2,\tilde{\epsilon}_3)=(--+), $  $(\tilde{h}_1,\tilde{h}_2,
\tilde{h}_3)=(h_3,h_1,h_2),$ and
\be
\phi_{\pm}(v)=\left(\frac{\sinh(v \pm i u)}{\sin \g}\right)^{N/2}\, ,\ u=J\sin\g \beta/N\, .
\ee
The $q_j(v)$ functions are defined as
\be
q_j(v)=\left\{ \begin{array}{lr} \phi_-(v)\, , & j=0\, ,\\
                                \prod_{k=1}^{N/2} \sinh(v-v_k^{(j)})\, ,&\quad j=1,2\, , \\
                                  \phi_+(v)\, , & j=3\, ,
               \end{array}\right.
\ee
with $\{v_k^{(1)}\}_{k=1}^{N/2}$, $\{v_k^{(2)}\}_{k=1}^{N/2}$  parameters which are called Bethe roots and
satisfy the quantum transfer matrix BAEs (see below). If we  introduce two auxiliary functions
\begin{subequations}\label{aux}
\begin{align}
\mathfrak{a}_1&=\frac{\lambda_1(v)}{\lambda_2(v)}=\frac{\phi_-(v+i\g)}{\phi_-(v)}\frac{q_1(v-i\g)}
{q_1(v+i\g)}\frac{q_2(v)}{q_2(v-i\g)}e^{\beta(h_3-h_1)}\, ,\\
\mathfrak{a}_2&=\frac{\lambda_3(v)}{\lambda_2(v)}=\frac{\phi_+(v+i\g)}{\phi_+(v)}
\frac{q_1(v)}{q_1(v+i\g)}e^{\beta(h_2-h_1)}\, ,
\end{align}
\end{subequations}
the BAEs of the quantum transfer matrix can be written as $(j=1,2)$:  $\mathfrak{a}_j(v_k^{(j)})=-1\, ,k=1,
\cdots,N/2 .$

\subsection{Integral equations for the auxiliary functions}

\begin{figure}
\includegraphics[width=\linewidth]{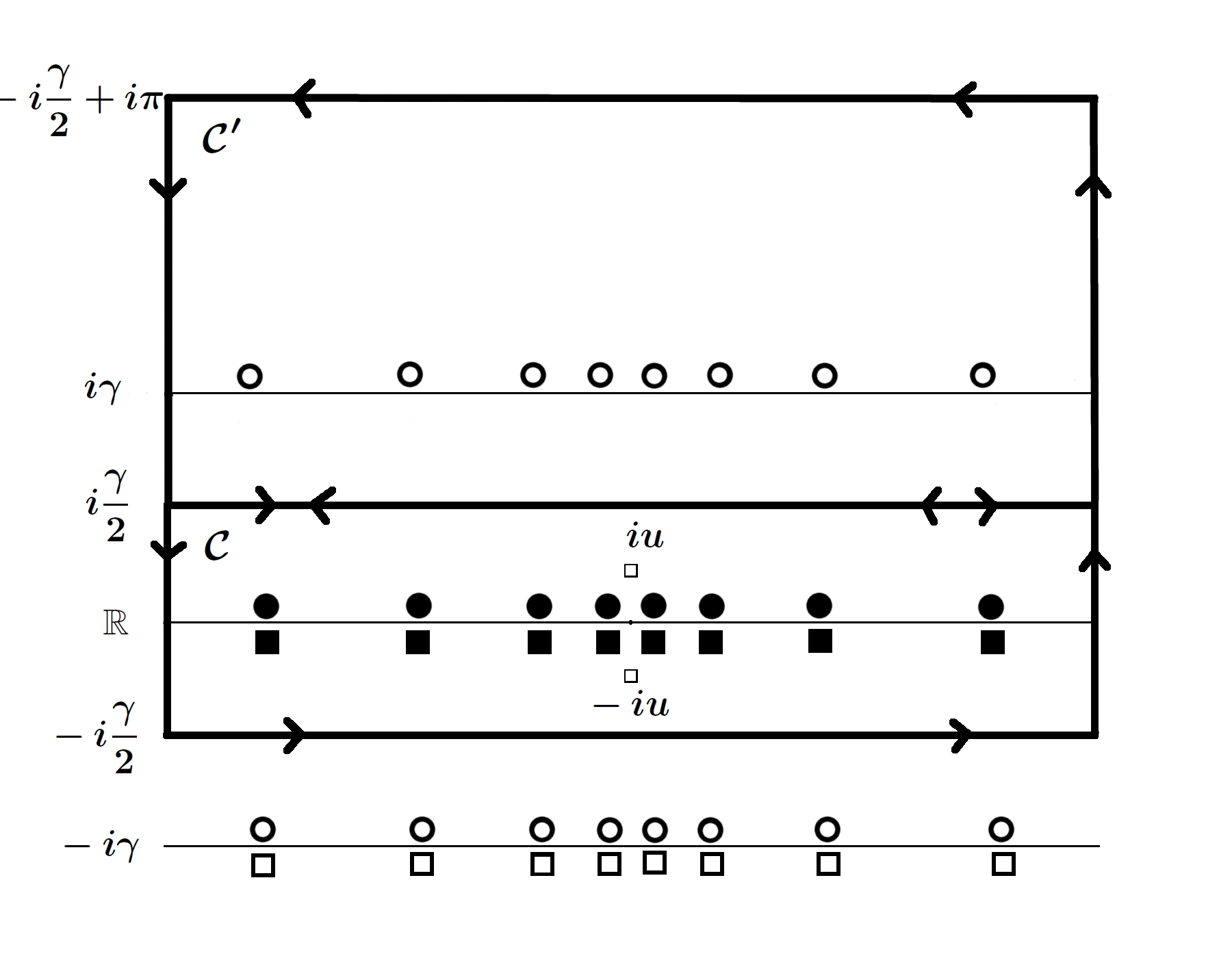}
\caption{Distribution of Bethe roots ($\blacksquare, \bullet$) and holes $(\square,\circ)$ for the largest eigenvalue of the QTM and  $\g\in(0,\pi/2)$. The contour
$\mathcal{C}$ contains all the Bethe roots and the poles of order $N/2$ at $\pm i u$. The lower edge of the
contour $\mathcal{C}'$ coincides with the upper edge of $\mathcal{C}$ but it has opposite orientation.
}
\label{contours}
\end{figure}

First, we will derive a set of NLIEs for the auxiliary functions
(\ref{aux}). Both of the functions are periodic of period $i\pi$. The equation
$\mathfrak{a}_1(v)=-1$ has $3N/2$ solutions, of which $N/2$ are the so-called
Bethe roots, $\{v_j^{(1)}\}_{j=1}^{N/2},$ and $N$ solutions, which are called
holes, and they are denoted by $\{{v'}_j^{(1)}\}_{j=1}^{N}$. However, the
second equation $\mathfrak{a}_2(v)=-1$ has only $N$ solutions, of which $N/2$
are the Bethe roots, $\{v_j^{(2)}\}_{j=1}^{N},$ and the other $N/2$ are the
second set of holes denoted by $\{{v'}_j^{(2)}\}_{j=1}^{N/2}$. A typical
distribution of Bethe roots and holes characterizing the largest eigenvalue of
the QTM for $\g\in(0,\pi/2)$ is shown in Fig.~\ref{contours}.
For any value of the Trotter number $N$ the strip $|\mbox{Im} v|<\g/2$
contains all the Bethe roots and the poles of order $N/2$ at $\pm
iu$. Introducing the rectangular contour $\mathcal{C}$ centered at the origin,
which extends to infinity and is depicted in Fig.~\ref{contours} we can define
for $v$ outside of $\mathcal{C}$ $(j=1,2)$
\begin{align}
f_j(v)&=\frac{1}{2\pi i}\int_{\mathcal{C}}\frac{d}{d v}[\ln\sinh(v-w)]\ln [1+\mathfrak{a}_j(w)]\,  dw\, , \nonumber\\
&=\frac{1}{2\pi i}\int_{\mathcal{C}}\ln \sinh(v-w)\frac{\mathfrak{a}_j'(w)}{1+\mathfrak{a}_j(w)}\,  dw\, .
\end{align}
The last relation was derived using integration by parts and the fact that the winding number of
$\ln [1+\mathfrak{a}_j(w)]$ is zero due to the fact that the number of zeroes and order of the poles inside the contour
is the same. Then, we can show that (see Sec. 6.3 of \cite{WW} or \cite{PK1,PK2})
\begin{subequations}\label{ii3}
\begin{align}
f_1(v)&=\ln q_1(v)-\ln \phi_-(v)-\frac{N}{2}\ln \sin \g\, ,\\
f_1(v)&=\ln q_2(v)-\ln \phi_+(v)-\frac{N}{2}\ln \sin \g\, .
\end{align}
\end{subequations}
Taking the logarithm of the auxiliary functions (\ref{aux}) and using the previous result (\ref{ii3}) we obtain
\begin{subequations}
\begin{align*}
\ln \mathfrak{a}_1(v)&=\beta(h_3-h_1)+\ln \left[\frac{\phi_+(v)}{\phi_-(v)}
\frac{\phi_-(v-i\g)}{\phi_+(v-i\g)}\nonumber\right]\\
&\ \  \ \ +f_1(v-i\g)-f_1(v+i\g)+f_2(v)-f_2(v-i\g)\, ,\\
\ln \mathfrak{a}_2(v)&=\beta(h_2-h_1)+\ln \left[\frac{\phi_-(v)}{\phi_+(v)}
\frac{\phi_+(v+i\g)}{\phi_-(v+i\g)}\nonumber\right]\\
&\ \  \ \ \ \ \ \ +f_1(v)-f_1(v+i\g)\, .
\end{align*}
\end{subequations}
Now we can take the Trotter limit, $\lim_{N\rightarrow\infty} \ln\left[\phi_+(v)/\phi_-(v)\right]
=i J \beta \sin\g\coth v$, with the result
\begin{subequations}\label{eq}
\begin{align}
\ln \mathfrak{a}_1(v)&=\beta(h_3-h_1)-\beta\frac{J\sinh^2 i\g}{\sinh v\sinh(v-i\g)}\nonumber\\
&\ \ \ \ \ \ \ \ \   +\int_{\mathcal{C}}\bar{K}_0(v-w)\ln[1+\mathfrak{a}_1(w)]\, dw\nonumber\\
&\ \ \ \ \ \ \ \ \  -\int_{\mathcal{C}}\bar{K}_2(v-w)\ln[1+\mathfrak{a}_2(w)]\,  dw\, ,\label{e1}\\
\ln \mathfrak{a}_2(v)&=\beta(h_2-h_1)-\beta\frac{J\sinh^2 i\g}{\sinh v\sinh(v+i\g)}\nonumber\\
&\ \ \ \ \ \ \  +\int_{\mathcal{C}}\bar{K}_1(v-w)\ln[1+\mathfrak{a}_1(w)]\,  dw\, ,\label{e2}
\end{align}
\end{subequations}
where
\begin{align}
\bar{K}_0(v)&=\frac{1}{2\pi i}\frac{\sinh 2i\g}{\sinh(v+i\g)\sinh(v-i\g)}\, ,\\
\bar{K}_1(v)&=\frac{1}{2\pi i}\frac{\sinh i\g}{\sinh(v)\sinh(v+i\g)}\, ,\\
\bar{K}_2(v)&=\frac{1}{2\pi i}\frac{\sinh i\g}{\sinh(v)\sinh(v-i\g)}\, .
\end{align}
Eqs.~(\ref{eq}) were derived assuming $\gamma\in(0,\pi/2)$ and $v$ is outside the contour. For $v$ inside
the contour  we need to add a $\ln[1+\mathfrak{a}_2(v)]$ term on the right hand side of Eq.~(\ref{e1}) and
a  $\ln[1+\mathfrak{a}_1(v)]$ term on the right hand side of Eq.~(\ref{e2}). For $\gamma\in(\pi/2,\pi)$ the
same equations remain valid if we replace $\mathcal{C}$ with a similar rectangular contour with horizontal
edges situated  at $\pm i(\pi-\g-\epsilon)/2$.

\subsection{Integral expression for the largest eigenvalue}\label{le}

The largest eigenvalue of the QTM is analytic in a strip around the real axis, therefore it will be
sufficient to derive an integral expression for  $\ln \Lambda_0(v_0)$ with $v_0$ close to the real axis
and then take the limit $v_0 \rightarrow 0$ to obtain the free energy. For our purposes we choose $v_0= iu$
for which $\lambda_3(v_0)=0$ and ($c$ is a constant)
\be\label{lei}
\Lambda_0(v_0)=\lambda_1(v_0)+\lambda_2(v_0)=c\, \frac{\phi_+(v_0)q_1^{(h)}(v_0)}{q_2(v_0)}\, ,
\ee
where we have used the identity (\ref{iden1}) and  $q_i^{(h)}(v)$ are defined in Appendix \ref{a2}.

Consider $v$ inside the contour $\mathcal{C}$. Then, inside the contour $\mathcal{C}'$ depicted in
Fig.~\ref{contours}, the function $1+\mathfrak{a}_1(v)$ has $N$ zeroes identified with the holes
$\{{v'}_j^{(1)}\}_{j=1}^N$, $N/2$ poles located at $\{v_j^{(1)}-i\g\}_{j=1}^{N/2}$ and $N/2$ poles located
at $\{v_j^{(2)}+i\g\}_{j=1}^{N/2}$ (some of the holes and poles are modulo $i\pi$). This means that around
$\mathcal{C}'$ the function  $\ln[1+\mathfrak{a}_1(v)]$ has zero winding number. Using the identity
(\ref{iden3}) in the form ($d(v)=d \ln\sinh v/dv$)
\[
\int_{\mathcal{C}}d(v-w)\frac{\mathfrak{a}_j'(w)}{1+\mathfrak{a}_j(w)}\, dw=
-\int_{\mathcal{C}'}d(v-w)\frac{\mathfrak{a}_j'(w)}{1+\mathfrak{a}_j(w)}\, dw
\, ,
\]
the right hand side can be computed as (\ref{ii3}) with the result
\begin{align}
&\frac{1}{2\pi i}\int_{\mathcal{C}}d(v-w)\frac{\mathfrak{a}_1'(w)}{1+\mathfrak{a}_1(w)}\, dw =\sum_{j=1}^{N/2} d(v-v_j^{(1)}+i\g)\nonumber\\
&\ \ \ \ \ \ \ \ \ +\sum_{j=1}^{N/2} d(v-v_j^{(2)}-i\g)-\sum_{j=1}^N d(v-{v'}_j^{(1)}).\,
\end{align}
After integration by parts with respect to $w$ and then integration with respect to $v$ we find
\begin{align}\label{i5}
&\frac{1}{2\pi i}\int_{\mathcal{C}}d(v-w)\ln[1+\mathfrak{a}_1(w)]\, dw =-\ln q_1^{(h)}(v)\nonumber\\
&\ \ \ \ \ \ \ \ \ +\ln q_1(v+i\g)+\ln q_2(v-i\g)+ c.\,
\end{align}
In a similar fashion using the fact that inside $\mathcal{C}'$ the function $1+\mathfrak{a}_2(v)$ has $N/2$ zeroes
at the holes $\{{v'}_j^{(2)}\}_{j=1}^{N/2}$ and $N/2$ poles located at$\{v_j^{(1)}-i\g\}_{j=1}^{N/2}$ (some modulo $i\pi$)
we find
\begin{align}\label{i6}
&\frac{1}{2\pi i}\int_{\mathcal{C}}d(v-w)\ln[1+\mathfrak{a}_2(w)]\, dw =-\ln q_2^{(h)}(v)\nonumber\\
&\ \ \ \ \ \ \ \ \ \ \ \ \ \ \ \ \ \ \ \ \ \ \ \ \ \ \ \ \ \ +\ln q_1(v+i\g)+ c.\,
\end{align}

For $v$ inside $\mathcal{C},$  $v\pm i\g$ is outside of the contour. Therefore, from (\ref{ii3})
we have
\begin{subequations}
\begin{align}
\frac{1}{2\pi i}&\int_{\mathcal{C}}d(v-w)\ln[1+\mathfrak{a}_1(w)]\, dw =\ln q_1(v+i\g)\label{i7a}\nonumber\\
&\ \ \ \ \ \ \ \ \ \ \ \ \ \ \ \ \ \ \ \ \ \ \ -\ln\phi_-(v+i\g)-\frac{N}{2}\sin\g\, ,\\
\frac{1}{2\pi i}&\int_{\mathcal{C}}d(v-w)\ln[1+\mathfrak{a}_2(w)]\, dw =\ln q_2(v-i\g)\label{i7b}\nonumber\\
&\ \ \ \ \ \ \ \ \ \ \ \ \ \ \ \ \ \ \ \ \ \ \ -\ln\phi_+(v-i\g)-\frac{N}{2}\sin\g\, .
\end{align}
\end{subequations}
Subtracting Eq.~(\ref{i5}) from Eq.(\ref{i7a}) and Eq.~(\ref{i6}) from Eq.~(\ref{i7b}) we obtain
\begin{subequations}\label{i8}
\begin{align}
\int_{\mathcal{C}} &\bar{K}_1(v-w)\ln[1+\mathfrak{a}_1(w)]\, dw=-\ln q_1^{(h)}(v)\nonumber\\
&\ \ \ \ \ \ \ +\ln q_2(v-i\g)+\ln\phi_-(v+i\g)+c\, ,\\
-\int_{\mathcal{C}} &\bar{K}_2(v-w)\ln[1+\mathfrak{a}_2(w)]\, dw=-\ln q_2^{(h)}(v)\nonumber\\
&\ \ \ \ \ \ \ \ \ \ \ \ \ +\ln q_1(v+i\g)-\ln q_2(v-i\g)\nonumber\\
& \ \ \ \ \ \ \ \ \ \ \ \ \ +\ln\phi_+(v-i\g)+c\, .
\end{align}
\end{subequations}
The importance of this result comes to light by noticing that the expression of the largest eigenvalue
(\ref{lei}) can be rewritten using (\ref{idena}) as
\begin{align*}
\ln\Lambda_0(v_0)&=\ln q_1^{(h)}(v_0)+\ln q_2^{(h)}(v_0)\nonumber\\
&\ \ \ \ \ \ \ \ \ \ -\ln q_1(v_0+i\g)-\ln[1+\mathfrak{a}_2(v_0)]+c\, ,
\end{align*}
and then using (\ref{i8}) as
\begin{align}\label{const}
&\ln\Lambda_0(v_0)=-\int_{\mathcal{C}} \bar{K}_1(v_0-w)\ln[1+\mathfrak{a}_1(w)]\,  dw\nonumber\\
&\ \ \ +\int_{\mathcal{C}} \bar{K}_2(v_0-w)\ln[1+\mathfrak{a}_2(w)]\, dw-\ln[1+\mathfrak{a}_2(v_0)]\nonumber\\
& \ \ \ +\ln[\phi_+(v_0-i\g)\phi_-(v_0+i\g)]+c\, .
\end{align}
The constant of integration can be computed by noticing that Eq.~(\ref{const}) is in fact valid for
all $v$ in a narrow strip around the real axis with $\ln[\lambda_1(v)+\lambda_2(v)]$ replacing the left
hand side. Considering the limit  $v \rightarrow \infty$ and using $\lim_{ v\rightarrow \infty}
[\lambda_1(v)+\lambda_2(v)]/[\phi_+(v-i\g)\phi_-(v+i\g)]=e^{\beta h_1}+e^{\beta h_3}$
we find
\be
c=\beta h_1+c'\, ,\   \mbox{ with } c'= 2 \ln[(1+e^{\beta(h_3-h_1)})(1+e^{\beta(h_2-h_1)})]
\ee
Finally, by taking
the Trotter limit, $N\rightarrow \infty,$ and  using $\lim_{ N\rightarrow\infty}
\ln[\phi_+(v_0-i\g)\phi_-(v_0+i\g)]=-J\cos\g\beta$ we obtain
\begin{align}\label{lelattice}
&\ln\Lambda_0(0)=c-J\beta \cos\g-\int_{\mathcal{C}} \bar{K}_2(w)\ln[1+\mathfrak{a}_1(w)]\,  dw\nonumber\\
&\ \ \ +\int_{\mathcal{C}} \bar{K}_1(w)\ln[1+\mathfrak{a}_2(w)]\, dw-\ln[1+\mathfrak{a}_2(0)]\, .
\end{align}
This result was derived for $\g \in(0,\pi/2)$ but it remains valid also for $\g\in(\pi/2,\pi)$ if $\mathcal{C}$
is replaced by a rectangular contour with the horizontal edges situated at $\pm i(\pi-\g-\epsilon)/2$.

\subsection{Continuum limit}

The continuum limit (see Sec. \ref{S6}) of the integral equations (\ref{eq}) and integral expression for the
largest eigenvalue  (\ref{lelattice}) is the same as the one performed for the 2CBG and is presented in detail
in \cite{PK1}. In the scaling limit we obtain Eq.~(\ref{grandQTM}) for the grandcanonical potential of the
continuum model with the auxiliary functions satisfying the NLIEs (\ref{nlieQTM}).

\section{Conclusions}\label{S8}

In this paper we have derived an alternative thermodynamic description for the
Bose-Fermi mixture in the QTM framework and performed a detailed analysis of
the contact at zero and finite temperature. In the strong coupling regime the
contact develops a pronounced local minimum as a function of the temperature
which is accompanied by a significant momentum reconstruction at both low and
large momenta. This momentum reconstruction can be experimentally detected and
provides an identification of the transition from the TLL to the
spin-incoherent regime. In addition, we have also showed that the boundaries
of the QC regions can be well mapped by the maxima of the grand canonical
specific heat. Our results also hint at the possibility of deriving efficient
thermodynamic descriptions for integrable $\kappa$-component ($\kappa>2)$,
systems with contact interactions involving only $\kappa$ integral equations.

\acknowledgments

O.I.P. acknowledges financial support from the LAPLAS 4 and 5 programs of the
Romanian National Authority for Scientific Research (CNCS-UEFISCDI).  Both
authors are grateful to Deutsche Forschungsgemeinschaft (DFG) for support via
Research Unit FOR 2316.

\appendix

\section{Some useful identities}\label{a2}

In this Appendix we prove certain identities which are needed in the derivation of the integral expression
of the largest QTM eigenvalue. First, we will prove that
\be\label{iden1}
\lambda_1(v)+\lambda_2(v)=c \, \frac{\phi_+(v)q_1^{(h)}(v)}{q_2(v)}\, ,
\ee
with $c$ a constant and $q_1^{(h)}(v)$ defined by
\be
q_1^{(h)}=\prod_{i=1}^{N}\sinh(v-{v'}_i^{(1)})\, .
\ee
From the definition of the $\lambda_j(v)$ functions we obtain
\[
\lambda_1(v)+\lambda_2(v)=\frac{\phi_+(v) p_1(v)}{q_1(v)q_2(v)}\, ,
\]
with $ p_1(v)=(\phi_-(v+i\g)q_1(v-i\g)q_2(v)e^{\beta
  h_3}$$+\phi_-(v)q_1(v+i\g)q_2(v-i\g)e^{\beta h_1})\, .  $ The equation
$p_1(v)=0$ (which is equivalent to $\mathfrak{a}_1(v)=-1$) has $3N/2$
solutions which are the $N/2$ Bethe roots, $\{v_j^{(1)}\}_{j=1}^{N/2},$ and
the $N$ holes $\{{v'}_j^{(1)}\}_{j=1}^N$. Also $p_1(v+i\pi)=(-1)^{3N/2}p(v)$
and $\lim_{v\rightarrow \infty}p_1(v)/\sinh ^{3N/2} v=const.$ which shows that
$p_1(v)=c \, q_1(v)q_1^{(h)}(v)$. This concludes the proof of (\ref{iden1}).

A similar identity is
\be\label{iden0}
\lambda_2(v)+\lambda_3(v)=c\,   \frac{\phi_-(v)q_2(v-i\g) q_2^{(h)}(v)}{q_1(v)}\, ,
\ee
with
\be
q_2^{(h)}=\prod_{i=1}^{N/2}\sinh(v-{v'}_i^{(2)})\, .
\ee
Again, from the definition we have
\[
\lambda_2(v)+\lambda_3(v)=\frac{\phi_-(v)q_2(v-i\g)p_2(v)}{q_1(v)q_2(v)}\, ,
\]
with
$
p_2(v)=\left(\phi_+(v)q_1(v+i\g)e^{\beta h_1}
 +q_1(v)\phi_+(v+i\g)e^{\beta h_2}\right)\, .
$
The equation $p_2(v)=0$ (equivalent to $\mathfrak{a}_2(v)=-1$) has $N$
solutions which are the $N/2$ Bethe roots, $\{v_j^{(2)}\}_{j=1}^{N/2}$, and
the $N/2$ holes $\{{v'}_j^{(2)}\}_{j=1}^{N/2}$.  In addition we have
$p_2(v+i\pi)=(-1)^N p_2(v)$ and $\lim_{v\rightarrow\infty} p_2(v)/\sinh^N
v=const.$ which shows that $p_2(v)$ can be written as $p_2(v)=c \,
q_2(v)q_2^{(h)}(v),$ concluding the proof of (\ref{iden0}). Also, we have
$\ln[1+\mathfrak{a}_2(v)]=\ln(p_2(v)/\phi_+(v)q_1(v+i\g)$ which is equivalent
to
\begin{align}\label{idena}
-&\ln\phi_+(v)+\ln q_2(v)-\ln q_1(v+i\g)\nonumber\\
& \ \   +\ln q_2^{(h)}(v) -\ln[1+\mathfrak{a}_2(v)]+const.=0\, .
\end{align}

In Sec. \ref{le} we will also use ($d(v)=\frac{d}{dv}\ln\sinh v$)
\be\label{iden3}
\int_{\mathcal{C}+\mathcal{C}'}d(v-w)\frac{\mathfrak{a}_j'(w)}{1+\mathfrak{a}_j(w)}\, dw=0\, ,
\ee
with the contours depicted in Fig.~\ref{contours}. The proof is similar with the one described
in \cite{PK1} and \cite{PK2} for the 2CBG and 2CFG cases and is left to the reader.

\end{document}